\documentclass[prc,twocolumn,showpacs,showkeys,floatfix]{revtex4}
\usepackage{graphicx}
\usepackage{subfigure}
\usepackage{color}
\usepackage{multirow}
\usepackage{setspace}
\usepackage{dcolumn}
\usepackage{amsmath}
\usepackage{float}
\usepackage{tablefootnote}
\usepackage{dcolumn}
\usepackage[justification=raggedright,labelfont=bf,singlelinecheck=false]{caption}
\usepackage{{bm}}

\definecolor{bluegray}{rgb}{0.4, 0.6, 0.8}
\usepackage{subfig}
\begin{document}
\title{Level Densities from 0-30 MeV}

\author{R.B.~Firestone}
\affiliation{University of California, Department of Nuclear Engineering,  Berkeley, CA 94720, USA}

\date{\today}

\begin{abstract}

Photon strength, $f(E_{\gamma})$, measured in photonuclear reactions, is the product of the average level density per MeV, $\rho(E_x)$, and the average reduced level width, $\Gamma_{\gamma}/E_{\gamma}^3$ for levels populated primarily by E1 transitions at an excitation energy $E_x=E_{\gamma}$.  It can be calculated with the Brink-Axel (BA) formulation modified to include contributions from the Giant Dipole Resonance (GDR) and higher lying resonances.  Level densities and reduced widths have been calculated for 17 nuclei with atomic numbers between Z=14-92.  Level densities below the GDR energy were calculated with the CT-JPI model and combined with the BA photon strength to determine the associated reduced widths.  The reduced widths varied exponentially with level energy and could be extrapolated up to higher energies.  The extrapolated widths were then combined with the BA photon strength to determine the level densities at higher energies.  The level densities are found to increase exponentially at low energies, peak near the GDR energy due to the appearance of new states at the $2\hbar\omega$ shell closure, and continue to increase less rapidly up to at least 30 MeV.  The average level densities have been compared with the Fermi Gas Level Density (FGLD), Back-Shifted Fermi Gas (BSFG), and Hartree-Fock-Bogoliubov (HFB) models.  Good agreement is found with the nearly identical FGLD and BDFG models, while the HFB models gives substantially lower level densities.  A universal set of FGLD model parameters were determined as a function of mass and temperature that are applicable to all nuclei.

\end{abstract}


\maketitle

\section{Introduction}

Photon strength for photonuclear reactions is defined as the product of the average level density per MeV, $\rho(E_x)$, and the average reduced level width, $\Gamma_{\gamma}/E_{\gamma}^3$, at the excitation energy $E_x=E_{\gamma}$ as shown in Eq.~\ref{GRS}.  It
\begin{equation}\label{GRS}
f(E_{\gamma}) = \rho (E_x) \cdot \Gamma_{\gamma}/E_{\gamma}^3
\end{equation}
peaks at the energy of the Giant Dipole Resonance (GDR) which is often interpreted as due to a collected enhancement of transition probability in which the protons in the nucleus move in one direction while the neutrons move in the opposite direction~\cite{Goldhaber46}.  There is no a priori evidence for such a collective behaviour nor is it observed in other physical systems.  As shown in Eq.~\ref{GRS} the GDR can be explained by either a sudden increase in the average level width or the average level density.  A sudden change in the level width is unexpected unless a new reaction channel were to open, but a sudden increase in level density is expected at each shell closure where a new ensemble of levels becomes available.  Indeed the energy of the GDR coincides with the energy of the $2\hbar\omega$ shell closure~\cite{Firestoneb21}.

\begin{figure}[!ht]
  \centering
  \includegraphics[width=8.5cm]{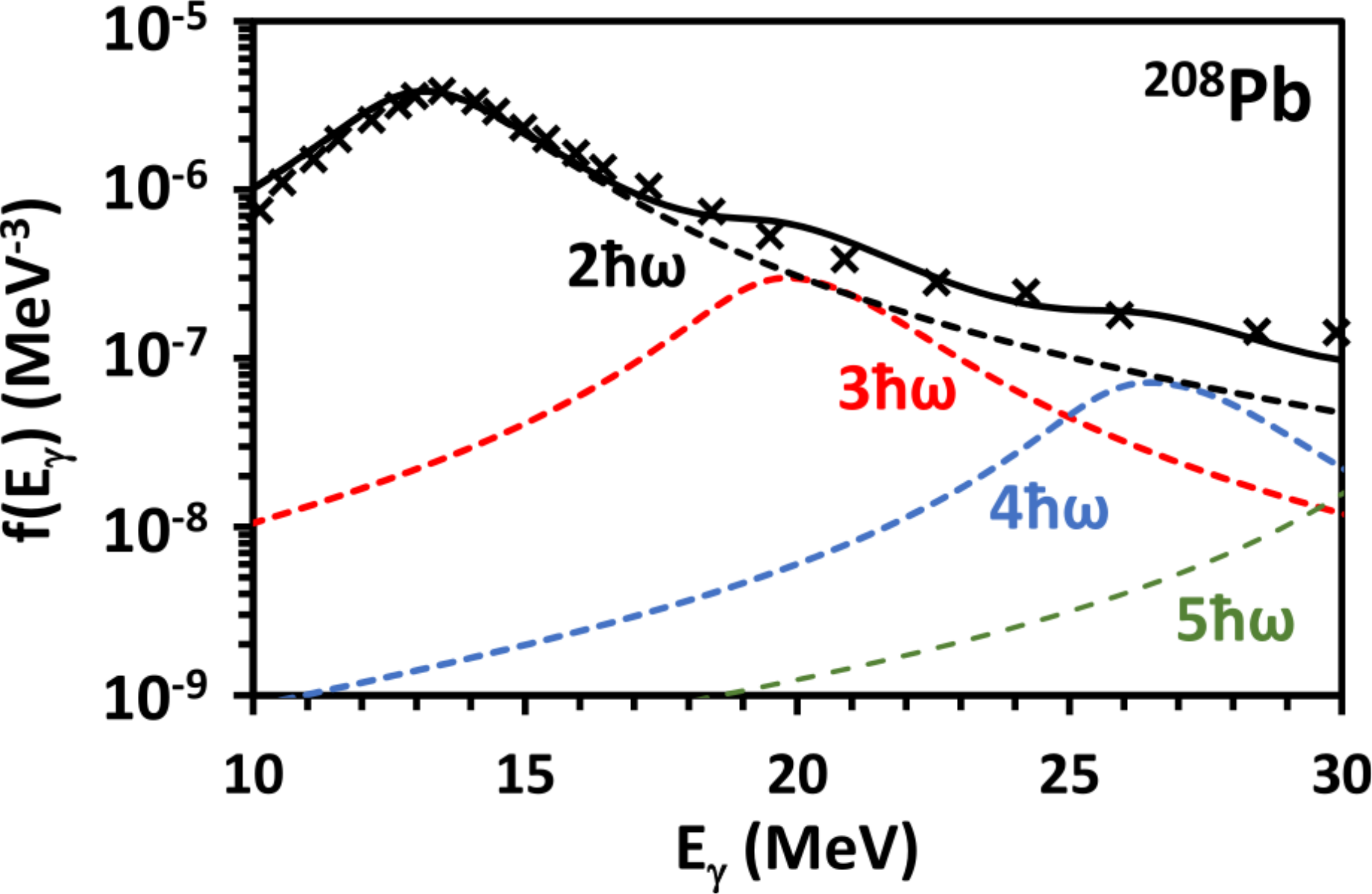}
  \caption{Calculated photon strength for the $^{208}$Pb($\gamma,n$) reaction for contributions from $2\hbar\omega-5\hbar\omega$ resonances.  The total calculated photon strength ($\boldsymbol{-}$) is compared with experiment (\textbf{x})~\cite{IAEA1178}.}
  \label{FS}
\end{figure}

The photon strength can be described by the Brink and Axel~\cite{Brink55,Axel62} formulation given Eq.~\ref{BA} where $E_i$ is the
\begin{equation}
f(E_{\gamma}) = \frac{1}{3(\pi \hbar c)^2}\sum_{i=1}^{i=2} \frac{\sigma_{i} E_{\gamma} \Gamma_{i}^2} {(E_{\gamma}^2-E_i^2)^2+E_{\gamma}^2\Gamma_i^2}
\label{BA}
\end{equation}
GDR energy in MeV,  $\Gamma_i$ is its width in MeV, and $\sigma_i$ is the photonuclear cross section in mb.  The summation is due to the splitting of GDR into two peaks in deformed nuclei leading to two sets of GDR peak parameters.  These parameters are determined by fitting them to experimental photonuclear data.  The BA formulation in Eq.~\ref{GRS} is incomplete because additional resonances will occur at higher shell closures that must be accounted for.  These resonances occur at $E_i=3\hbar\omega,~4\hbar\omega,...$ and contribute to the photon strength at all lower energies.  These contributions can also be calculated with the BA formulation where their relative cross sections, $\sigma_i$, can be determined from detailed balance as defined by Uhl and Kopecky~\cite{Uhl94} and given in Eq.~\ref{SG}.  The higher resonance widths are
\begin{equation}
\label{SG}
f(E_{\gamma}) = \frac{\sigma(E_{\gamma})} {3\pi^2 \hbar^2 c^2 E_{\gamma}}
\end{equation}
unknown but can be assumed to be the same as for the GDR which is sufficient to calculate their contributions to the photon strength at lower energies.  A comparison of the calculated photon strength with experiment for $^{208}$Pb~\cite{IAEA1178} is shown in Fig.~\ref{FS}.  The agreement between experiment and calculation is excellent, especially at higher energies where additional strength, previously attributed to multiple neutron emission, is consistent with the $(\gamma,n)$ reaction.

The BA formulation is unreliable at lower energies where strong pygmy (E1) and spin-flip (M1) transitions can dominate.  The contribution of M1 and E2 multipolarity transitions is small, $\lesssim$1\%~\cite{Kopecky90}.  The contributions of M1 and E2 transitions will be ignored here.  The calculated level density populated by photonuclear reactions is constrained by the E1 transition multipolarity to levels with $J^{\pi}=J_{GS}^{-\pi},J_{GS}^{-\pi}\pm1$.  For even-even nuclei this includes only $J^{\pi}=1^-$, and for nuclei with $J_{GS}=1/2^{+,-}$ only levels with $J^{\pi}=1/2^{-,+},3/2^{-,+}$ are populated.
\vspace{-1cm}
\section{Deconstructing the photon strength function}

The Brink-Axel photon strength function, as described in Eq.~\ref{BA}, requires six parameters, the GDR energies, $E_{1,2}$, widths, $\Gamma_{1,2}$, and cross sections, $\sigma_{1,2}$ for each contributing $(n\hbar)$ resonance.  These parameters have been measured for many nuclei and recommended values have been compiled~\cite{Kawano19}.  The parameters are also systematic and can be accurately described as a function of mass, $A$, and deformation, $\beta_2$~\cite{Firestone20}.   The GDR energies are calculated from their centroid energies , $\overline{E}(n\hbar\omega)$ and deformation by Eq.~\ref{EGDR},
\begin{equation}\label{EGDR}
\begin{aligned}
\overline{E}(n\hbar\omega) = &23.78(7)(A^{-1/3}-A^{-2/3})n\\
E_i(n\hbar\omega) = &\overline{E}(n\hbar\omega)\pm5.5(3)\beta_2
\end{aligned}
\end{equation}
widths by Eq.~\ref{Wid},
\begin{equation}\label{Wid}
\begin{aligned}
\Gamma_1(n\hbar\omega) =&7.41(15)A^{-1/6}\\
\Gamma_2(n\hbar\omega) =&11.13(16)A^{-1/6}
\end{aligned}
\end{equation}
and the total cross section by Eq.~\ref{S12}.
\begin{equation}\label{S12}
\begin{aligned}
\sigma_1(2\hbar\omega) + \sigma_2(2\hbar\omega) = &0.483(6)A^{4/3}\\
\sigma_2(2\hbar\omega) = &1.5\sigma_1(2\hbar\omega)\\
\sigma_i(n\hbar\omega) = &\sigma_i(2\hbar\omega)\frac{\overline{E}(n\hbar\omega)f[\overline{E}(n\hbar\omega)]}{\overline{E}(2\hbar\omega)f[\overline{}\overline{E}((2\hbar\omega)]}
\end{aligned}
\end{equation}
If the level density for states populated by photonuclear reactions, $\rho(E_x)$, is known then the corresponding average reduced widths, $\Gamma_{\gamma}/E_{\gamma}^3$ can be extracted from the photon strength using Eq.~\ref{GRS}.

\subsection{The CT-JPI level density model}

Level densities for all $J^{\pi}$ values can be fit by the CT-JPI model assuming a constant temperature, $T$, and separate energy cutoffs, $E_0 (J^{\pi})$ for each spin and parity~\cite{Firestoneb21}.  This differs from the conventional Constant Temperature (CT)~\cite{Eric60} and Back-Shifted Fermi Gas (BSFG)~\cite{Bethe36,Bethe37} models which incorrectly attempt to fit the total level density with the same two parameters.  The CT-JPI level densities are given by Eq.~\ref{CTJPI} where $\sigma_c$ is the spin cutoff
\begin{equation}\label{CTJPI}
\rho(E_x,J^{\pi}) = \textrm{exp}\bigg[\frac{E_x-E_0 (J^{\pi})}{T}\bigg])/T.
\end{equation}
parameter.  The $E_0 (J^{\pi})$, and $T$ parameters are fit to experimental level energies and $J^{\pi}$ sequences and they are constrained by the statistical spin distribution function~\cite{Gilbert65}, given by Eq.~\ref{SDF}.  The spin distribution function
\begin{equation}\label{SDF}
f(J) = \frac{2J+1}{2\sigma_c^2}\textrm{exp}\bigg[-\frac{(J+1/2)^2}{2\sigma_c^2}\bigg]
\end{equation}
is independent of parity so $E_0$, $T$, and $\sigma_c$ are least squares fit such that $f(J)=f(J^{\pi=+}) +f(J^{\pi=-})$ and the fractions of both parities vary smoothly with spin.  Notably the spin distribution function is independent of energy because all $\rho(E_x,J^{\pi})$ vary with the same temperature.  CT-JPI parameters have been evaluated for selected nuclei~\cite{Firestoneb21}, and they can also be estimated from the yrast energies and the systematic temperatures of nearby nuclei.

\subsection{Determination of reduced level widths}

\begin{figure}[!ht]
  \centering
  \includegraphics[width=8.5cm]{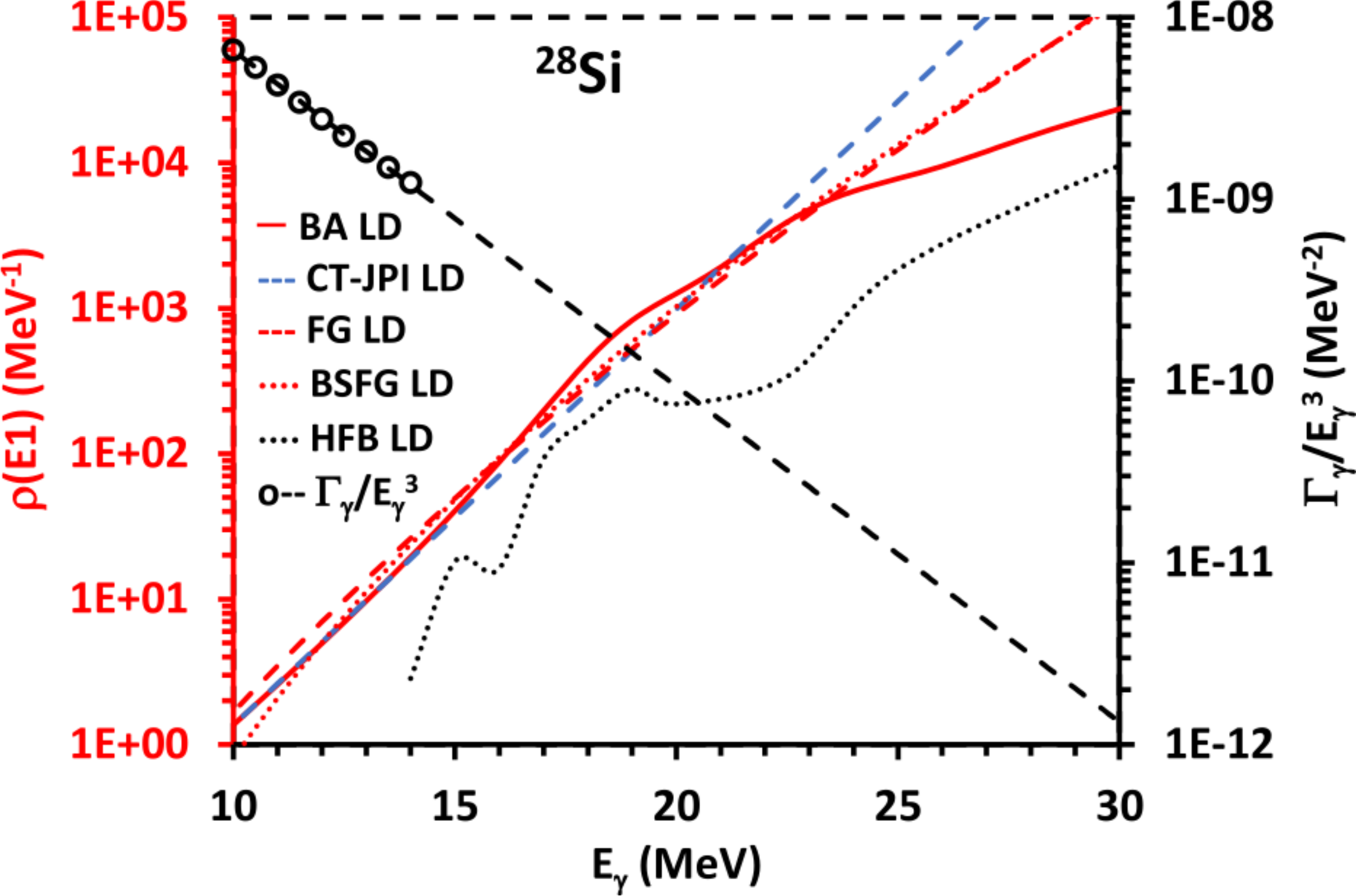}
  \caption{Average reduced widths calculated from CT-JPI and BA models below the GDR energy (\textbf{o}) and extrapolated to 30 MeV (\textbf{-~-~-}).  Level densities are from the CT-JPI model (\textcolor{bluegray}{\textbf{-~-~-}}), fitted to the BA model (\textcolor{red}{$\boldsymbol{-}$}), and calculated with the FGLD (\textcolor{red}{\textbf{-~-~-}}), BSFG (\textcolor{red}{\textbf{$\cdot\cdot\cdot$}}), and HFB (\textbf{$\cdot\cdot\cdot$}) models.}\label{28Si}
\end{figure}

The CT-JPI model is valid in at low energies where the contribution of the GDR is minor.  In this energy region the reduced $\gamma$-ray width can be determined by Eq.~\ref{GGE}.  The
\begin{equation}
\label{GGE}
\Gamma_{\gamma}/E_{\gamma}^3 = f(E_{\gamma})/\rho(E_x)_{CTJPI}
\end{equation}
low energy reduced $\gamma$-ray widths are found to vary smoothly and can be fit to an exponential equation of the form $(\Gamma_{\gamma}/E_{\gamma}^3)_{fit} =Ae^{BE_{\gamma}}$ where A and B are fitting constants.  The fitted reduced widths can then be used to calculated the level densities at higher energies by Eq.~\ref{RHO}.  The extrapolated level densities are valid assuming that
\begin{equation}
\label{RHO}
 \rho(E_x) = f(E_{\gamma})/Ae^{BE_{\gamma}}
\end{equation}
the exponential slope of the reduced width function doesn't change at higher energies.  They are also independent of the opening of new reaction channels because they are defined with respect to partial widths associated only with the BA photon strength function.
An example of this fitting procedure for $^{28}$Si is shown in Fig.~\ref{28Si}.  Here the BA parameters are given in Table~\ref{BAP} and the CT-JPI parameters are given in Table~\ref{TE0}.
The $^{28}$Si reduced widths derived from the CT-JPI and BA models vary exponentially with constants $A=1.07\times10^{-6}$ and $B=-0.539$.  The level density begins to increase with respect to the CT-JPI model at $\approx$15 MeV, has two peaks at the GDR energies, $\approx$19 MeV and $\approx$23 MeV, and then continues to increase more slowly than the CT-JPI model at higher energies.

\subsection{Fermi Gas level Density Model (FGLD)}

The Fermi Gas Level Ddensity (FGLD) model, as described by Sen’kov and Zelevinsky~\cite{Senkov16}, is given in Eq.~\ref{FGas}
\begin{equation}\label{FGas}
\begin{aligned}
\rho(E_{\gamma},M) &= \frac{N_{FG}}{E_{\gamma}^{5/4}}\textrm{exp}\bigg[2\sqrt{aE_{\gamma}}+c-\frac{M^2}{2\sigma^2}\bigg]\\
\sigma^2 &= \alpha\sqrt{E_{\gamma}}(1+\beta E_{\gamma})
\end{aligned}
\end{equation}
for $M$ spin projections where $\sigma$ is an energy dependent spin cutoff parameter and the level density parameter $a$ is the density of single-particle states at the Fermi surface.  Notably the spin projection term is small and makes only a minor contribution to the level density.  An arbitrary normalization parameter, $N_{FG}$, is added for comparison of the FGLD model and BA/CT-JPI model level densities.

The FGLD model level densities are fit to the BA/CT-JPI level densities in Fig.~\ref{28Si} using the parameters $\alpha\!=\!2.38$, $c\!=\!-2.97$, and $a\!=\!7.44$, with a normalization factor $N_{FG}\!=\!2\!\times\!10^{-5}$.  The parameters agree well with $\alpha\!=\!2.37(10)$ and $c\!=\!-2.92(13)$ reported by Sen’kov and Zelevinsky.  The agreement with the BA/CT-JPI level densities is good although the FGLD model fails to account for GDR level density peaks.

\renewcommand*{\arraystretch}{1.07}
\begin{table}[!h]
\small
\tabcolsep=2pt
\caption{\label{BAP} Modified BA GDR fitting parameters.}
\begin{tabular}{rlcccccc}
\toprule
\vspace{-0.3cm}\\
$^A$El$_Z$&$\beta_2^{~a}$&$\sigma_1$&$E_1$&$\Gamma_1$&$\sigma_2$&$E_2$&$\Gamma_2$\\
\colrule
\vspace{-0.3cm}\\
$^{238}$U$_{92}~2\hbar\omega$&0.289$^b$&285.5&11.28&2.98&428.3&14.46&4.47\\
$3\hbar\omega$&&74.5&16.92&2.98&111.8&21.69&4.47\\
$4\hbar\omega$&&21.8&22.55&2.98&32.7&28.93&4.47\\
$^{235}$U$_{92}~2\hbar\omega$&0.215&280.8&11.73&2.98&421.1&14.10&4.48\\
$3\hbar\omega$&&54.8&17.59&2.98&82.3&21.15&4.48\\
$4\hbar\omega$&&16.9&23.45&2.98&25.3&28.20&4.48\\
$^{208}$Pb$_{82}~2\hbar\omega$&0.054$^b$&238.6&13.04&3.04&357.9&13.64&4.57\\
$3\hbar\omega$&&29.0&19.56&3.04&43.5&20.46&4.57\\
$4\hbar\omega$&&9.6&26.08&3.04&14.5&27.28&4.57\\
$^{207}$Pb$_{82}~2\hbar\omega$&0.000&237.1&13.36&3.05&355.6&13.36&4.58\\
$3\hbar\omega$&&27.2&20.04&3.05&40.7&20.04&4.58\\
$4\hbar\omega$&&9.2&26.72&3.05&13.8&26.72&4.58\\
$^{164}$Ho$_{67}~2\hbar\omega$&0.284&173.8&12.63&3.17&260.7&15.77&4.76\\
$3\hbar\omega$&&38.7&18.95&3.17&58.1&23.65&4.76\\
$4\hbar\omega$&&11.5&25.27&3.17&17.2&31.54&4.76\\
$^{163}$Ho$_{66}~2\hbar\omega$&0.284&172.4&12.66&3.17&258.6&15.79&4.76\\
$3\hbar\omega$&&38.3&18.98&3.17&57.5&23.69&4.76\\
$4\hbar\omega$&&11.4&25.31&3.17&17.0&31.58&4.76\\
$^{164}$Dy$_{66}~2\hbar\omega$&0.296&173.8&12.57&3.17&260.7&15.83&4.76\\
$3\hbar\omega$&&40.5&18.85&3.17&60.8&23.75&4.76\\
$4\hbar\omega$&&11.9&25.13&3.17&17.9&31.67&4.76\\
$^{161}$Dy$_{66}~2\hbar\omega$&0.271&169.6&12.77&3.18&254.3&15.76&4.77\\
$3\hbar\omega$&&35.8&19.16&3.18&53.7&23.65&4.77\\
$4\hbar\omega$&&10.7&25.54&3.18&16.1&31.53&4.77\\
$^{133}$Ba$_{56}~2\hbar\omega$&0.151&131.4&14.15&3.28&197.1&15.81&4.93\\
$3\hbar\omega$&&18.0&21.22&3.28&27.0&23.72&4.93\\
$4\hbar\omega$&&5.7&28.30&3.28&8.6&31.63&4.93\\
$^{132}$Ba$_{56}~2\hbar\omega$&0.162&130.1&14.12&3.28&195.2&15.90&4.93\\
$3\hbar\omega$&&18.3&21.17&3.28&27.5&23.86&4.93\\
$4\hbar\omega$&&5.8&28.23&3.28&8.7&31.81&4.93\\
$^{116}$In$_{49}~2\hbar\omega$&0.146&109.5&14.70&3.36&164.3&16.31&5.04\\
$3\hbar\omega$&&14.4&22.04&3.36&21.6&24.46&5.04\\
$4\hbar\omega$&&4.6&29.39&3.36&6.9&32.62&5.04\\
$^{115}$In$_{49}~2\hbar\omega$&0.115&108.3&14.73&3.36&162.4&16.34&5.05\\
$3\hbar\omega$&&14.2&22.09&3.36&21.3&24.51&5.05\\
$4\hbar\omega$&&4.5&29.46&3.36&6.8&32.68&5.05\\
$^{72}$Ge$_{32}~2\hbar\omega$&0.241$^b$&58.0&16.04&3.63&87.0&18.69&5.46\\
$3\hbar\omega$&&8.8&24.05&3.63&13.3&28.04&5.46\\
$4\hbar\omega$&&2.7&32.07&3.63&4.1&37.39&5.46\\
$^{71}$Ge$_{32}~2\hbar\omega$&0.207&56.9&16.28&3.64&85.4&18.56&5.47\\
$3\hbar\omega$&&7.9&24.42&3.64&11.8&27.84&5.47\\
$4\hbar\omega$&&2.5&32.56&3.64&3.7&37.13&5.47\\
$^{44}$Ca$_{20}~2\hbar\omega$&0.253$^b$&37.6&17.91&3.94&45.1&20.70&5.92\\
$3\hbar\omega$&&4.9&26.87&3.94&6.4&26.87&5.92\\
$4\hbar\omega$&&1.5&35.82&3.94&2.0&35.82&5.92\\
$^{40}$Ca$_{20}~2\hbar\omega$&0.021&26.5&19.68&4.01&39.7&19.68&6.02\\
$3\hbar\omega$&&2.5&29.52&4.01&1.4&29.52&6.02\\
$4\hbar\omega$&&0.8&39.35&4.01&1.1&39.35&6.02\\
$^{28}$Si$_{14}~2\hbar\omega$&0.363&16.5&19.00&4.25&24.7&23.01&6.39\\
$3\hbar\omega$&&2.9&28.50&4.25&4.3&34.51&6.39\\
$4\hbar\omega$&&0.9&38.00&4.25&1.3&46.02&6.39\\
\botrule
\vspace{-0.3cm}\\
\multicolumn{8}{l}{$^a$Calculated values from except as noted~\cite{Moller16}.}\\
\multicolumn{8}{l}{$^b$From experimental B(E2) values~\cite{Pritychenko14}.}\\
\end{tabular}
\end{table}

The FGLD model is comparable to the Back-Shifted Fermi Gas (BSFG) model~\cite{Newton56} given by Eq.~\ref{BSFG} where $a$
\begin{equation}\label{BSFG}
\begin{aligned}
&\rho(E_x,J^{\pi}) = f(J^{\pi})\frac{\textrm{exp}\big[2\sqrt{a(E_{\gamma}-E_1)}~\big]}{12\sqrt{2}\sigma_ca^{1/4}(E_{\gamma}-E_1)^{5/4}}\\
&\sigma_c^2 = 0.0146A^{5/3}\frac{1+\sqrt{1+4a(E_{\gamma}-E_1)}}{2a}
\end{aligned}
\end{equation}
is a shell model level density parameter, $E_1$ is a backshift parameter, typically taken as the neutron separation energy, $\sigma_c$ is a spin cutoff parameter, and $f(J^{\pi})$ is a normalization factor for comparison with the BSFG model.  For $^{28}$Si a fit of the BSFG parameters to the FGLD level densities assuming a backshift, $E_1$=8.905 MeV, gives $a$=5.24, comparable to $a$=5.04 from RIPL-3~\cite{Capote09}, and $f(1^-)$=0.261.   The $^{28}$Si BSFG level densities plotted on Fig~\ref{28Si} are nearly identical to those calculated with the FGLD model.

\subsection{Hartree-Fock-Bogoliubov (HFB) calculation}

The RIPL-3 library~\cite{Capote09} provides an extensive set of Hartree-Fock-Bogoliubov (HFB) plus combinatorial nuclear level densities assuming ground state deformations for all nuclei and $J^{\pi}$ values.  The nuclear level density is coherently obtained on the basis of the single-particle level scheme and pairing energy derived at the ground state deformation based on the BSk14 Skyrme force~\cite{Goriely07,Goriely08}.  The HFB level densities for $^{28}$Si are shown in Fig.~\ref{28Si}.  These level densities are substantially lower than the other calculations.

\section{Results}

Level densities and average reduced widths calculated for $^{40,44}$Ca, $^{71,72}$Ge, $^{115,116}$In, $^{132,133}$Ba, $^{161,164}$Dy, $^{163,164}$Ho, $^{207,208}$Pb, and $^{235,238}$U are shown in Fig.~\ref{Plot}.  The low energy reduced level widths were fit to the exponential equation $(\Gamma_{\gamma}/E_{\gamma}^3)_{fit} =Ae^{BE_{\gamma}}$, as described above, and the level densities were calculated using the BA photon strength parameters listed in Table~\ref{BAP}.  The derived fitting parameters $A$ and $B$ are given in Table~\ref{TE0}.  In all cases the average reduced widths could be fit to an exponential with an uncertainty of $<$2\% and a coefficient of determination $R^2>0.994$.

The level densities in Fig.~\ref{Plot} are only for $J^{\pi}=1^-$ states for even-even nuclei, $J^{\pi}=1/2^-,3/2^-$ states for $J^{\pi}_{GS}=1/2^+$ nuclei, and $J^{\pi}=J_{GS}^{-\pi},J_{GS}^{-\pi}\pm1$ for all other nuclei.  The level densities all follow the CT-JPI model at low energies, peak near the GDR, and continue to increase more slowly than for the CT-JPI model to higher energies.  The peak in level density at the GDR is most pronounced at the doubly magic nuclei $^{40}$Ca and $^{208}$Pb where the total level density is low.  For deformed nuclei the GDR peak is washed out by the much higher level density.

\begin{figure*}[!t]
\centering
  \includegraphics[width=8.5cm]{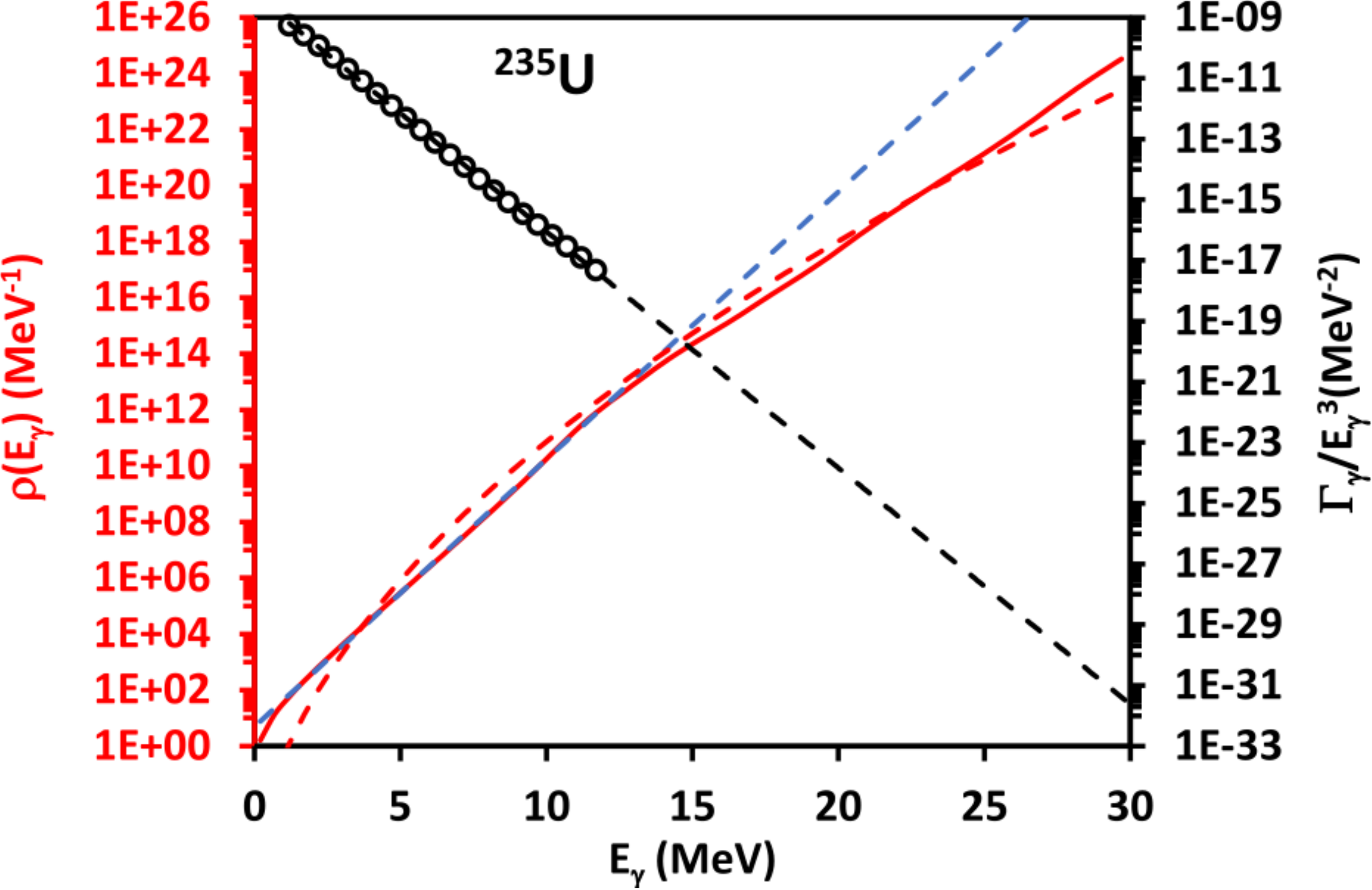}
  \includegraphics[width=8.5cm]{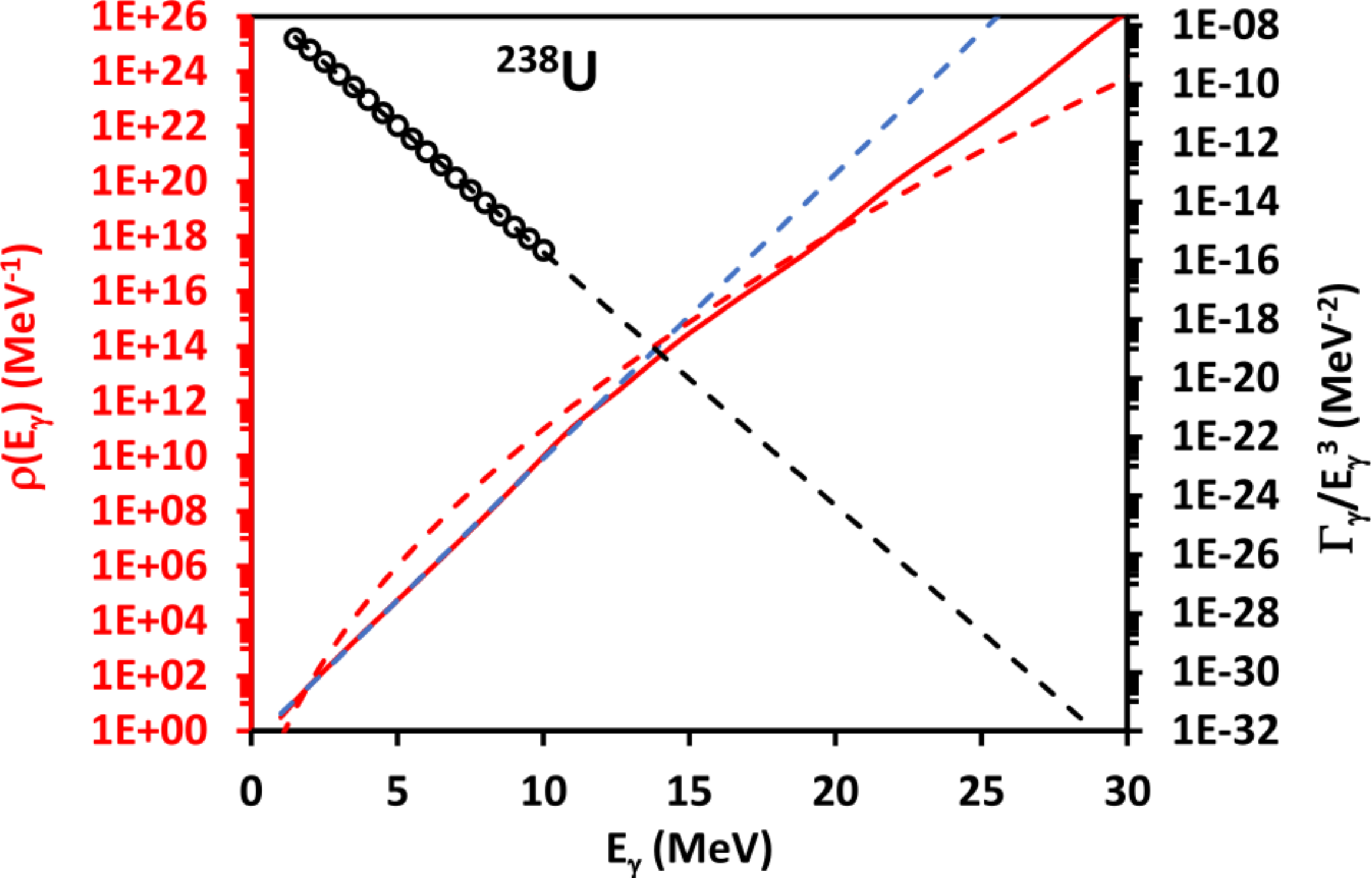}
\includegraphics[width=8.5cm]{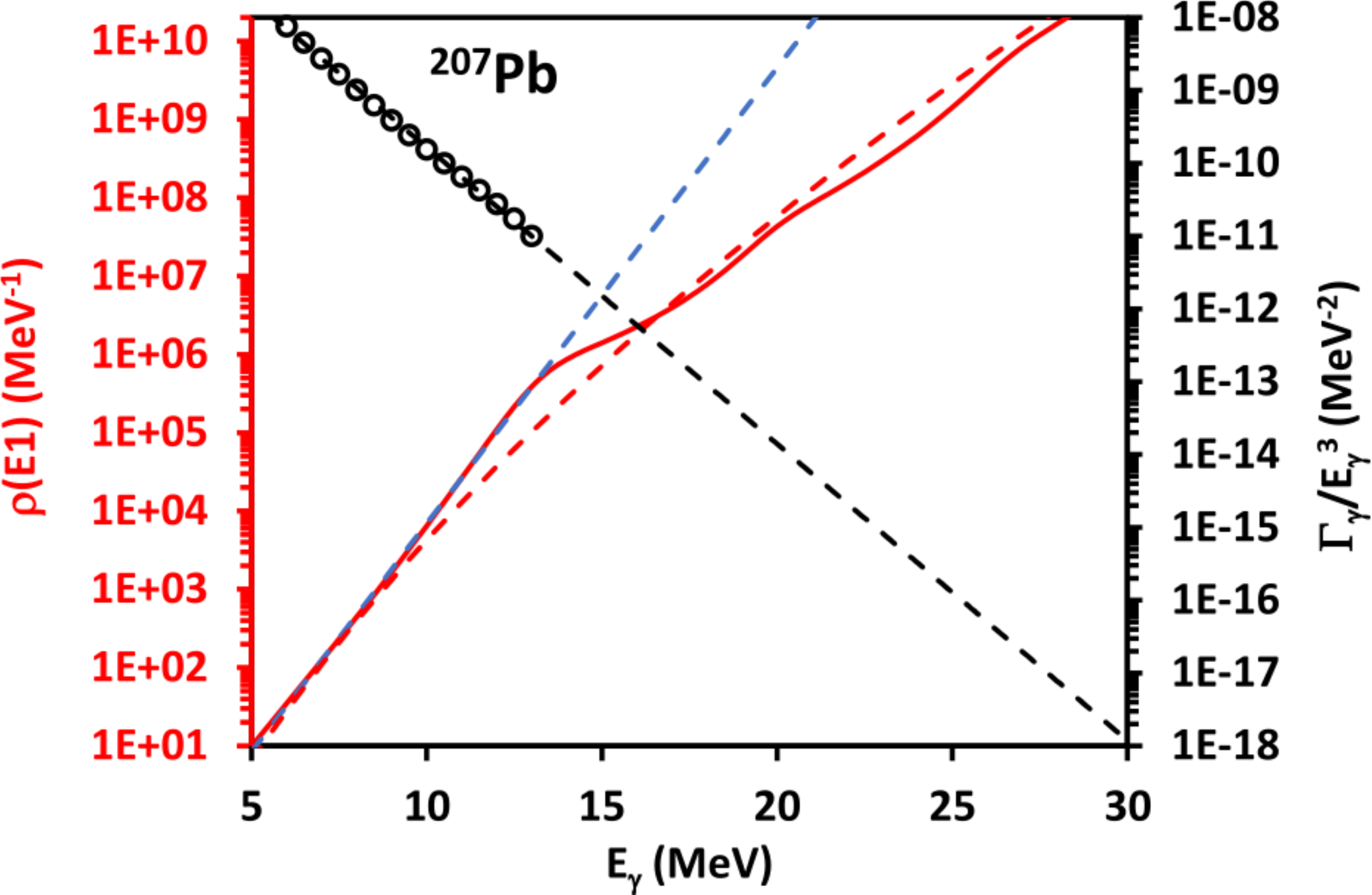}
\includegraphics[width=8.5cm]{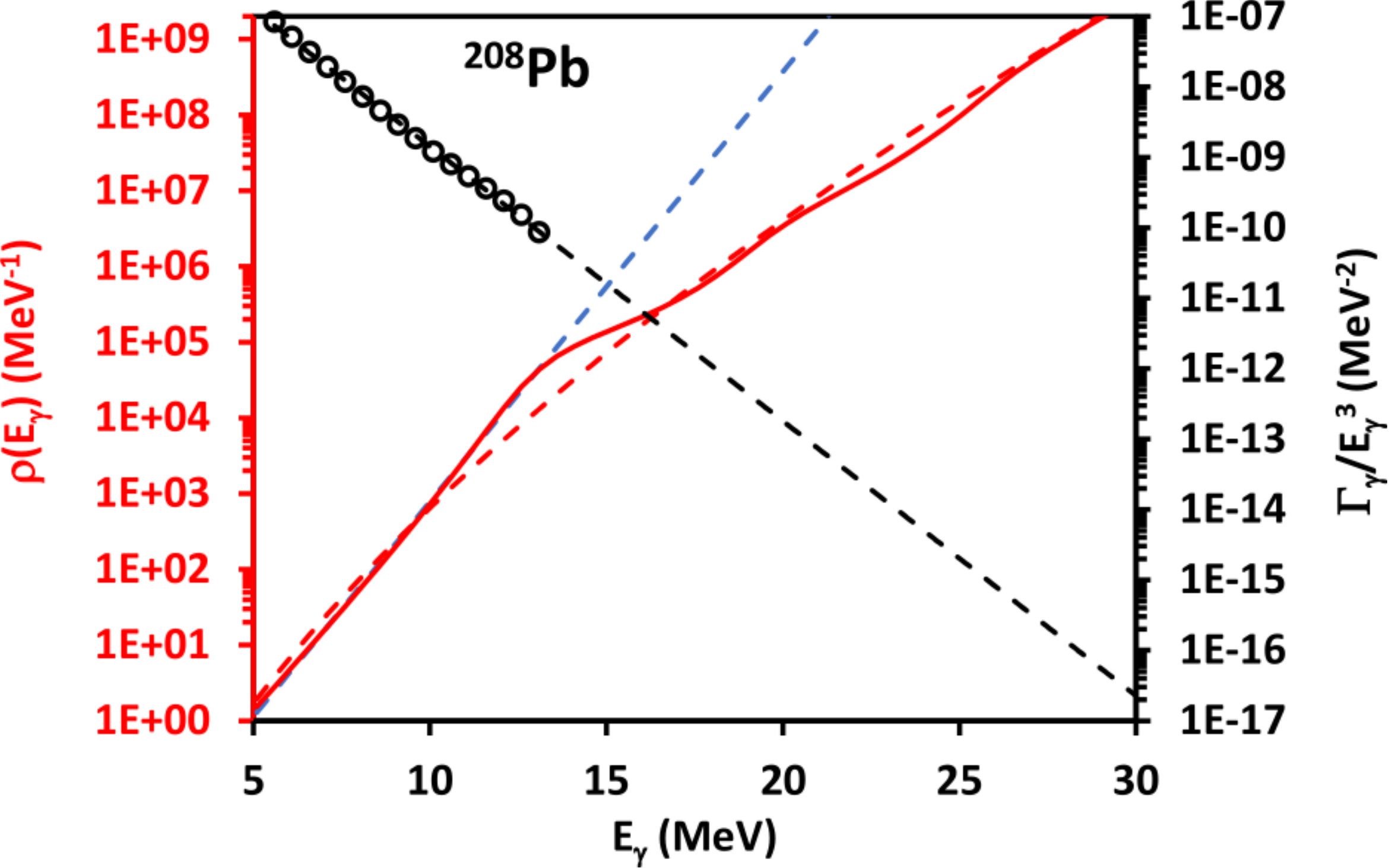}
\includegraphics[width=8.5cm]{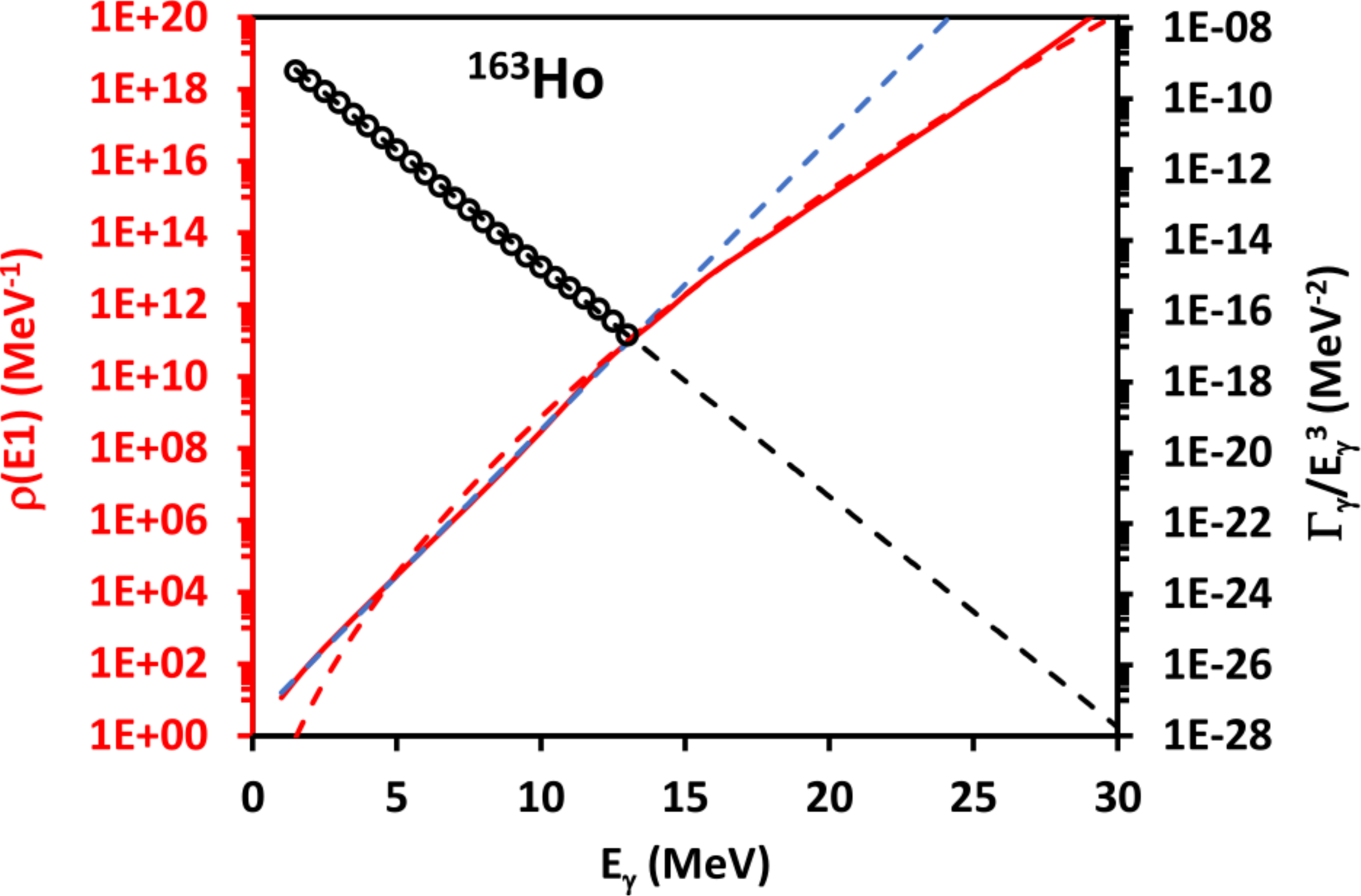}
\includegraphics[width=8.5cm]{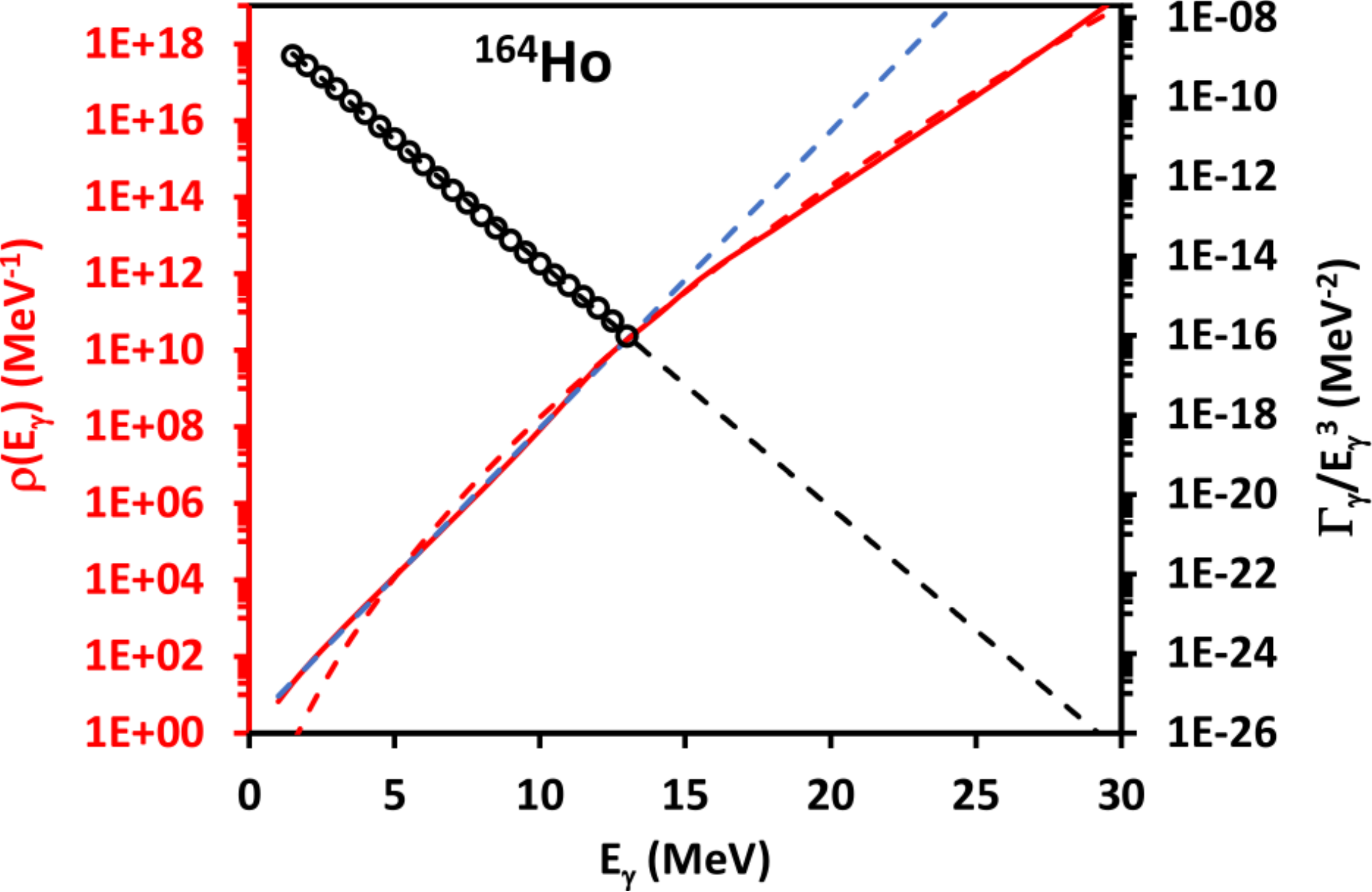}
\includegraphics[width=8.5cm]{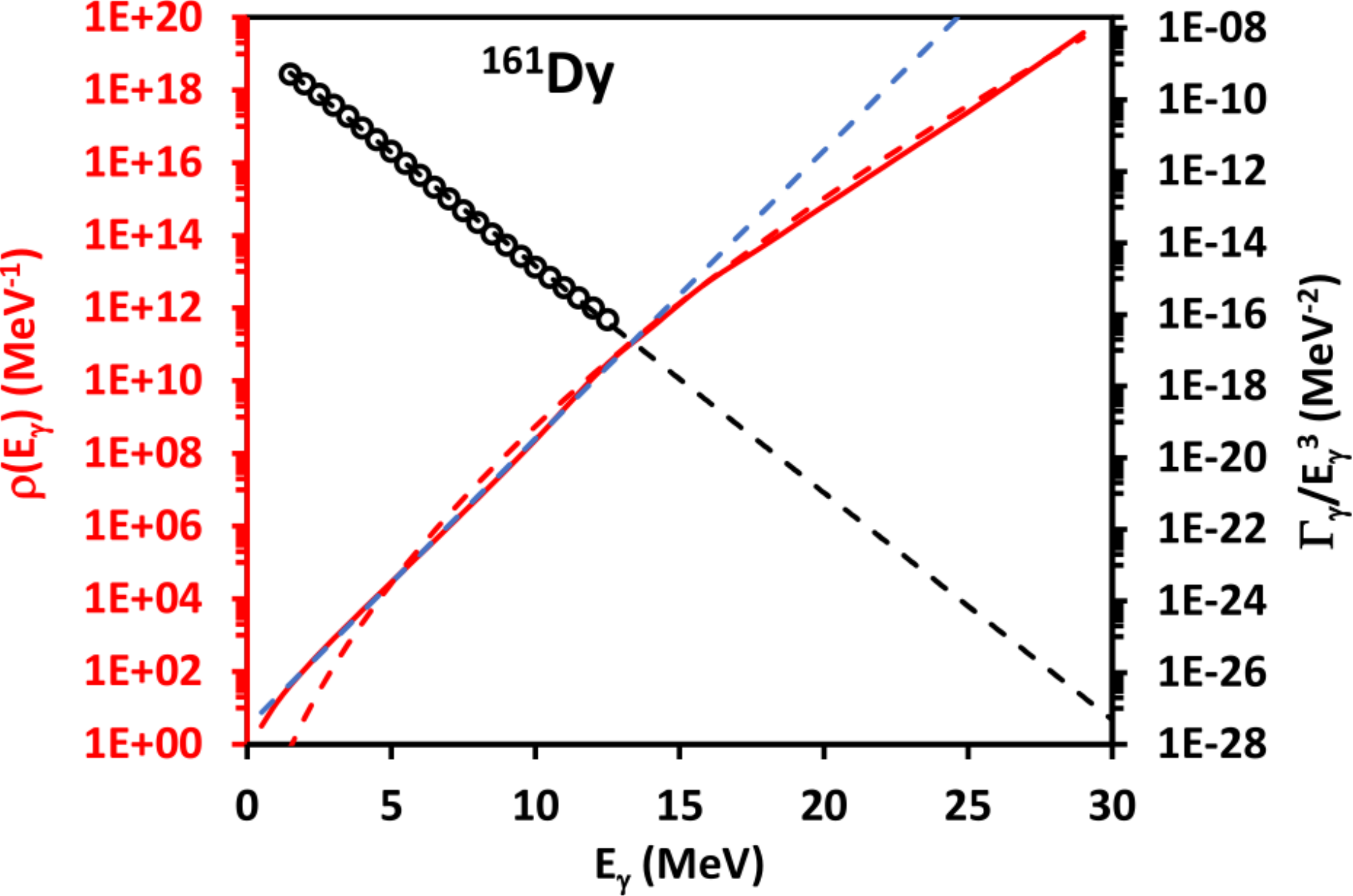}
\includegraphics[width=8.5cm]{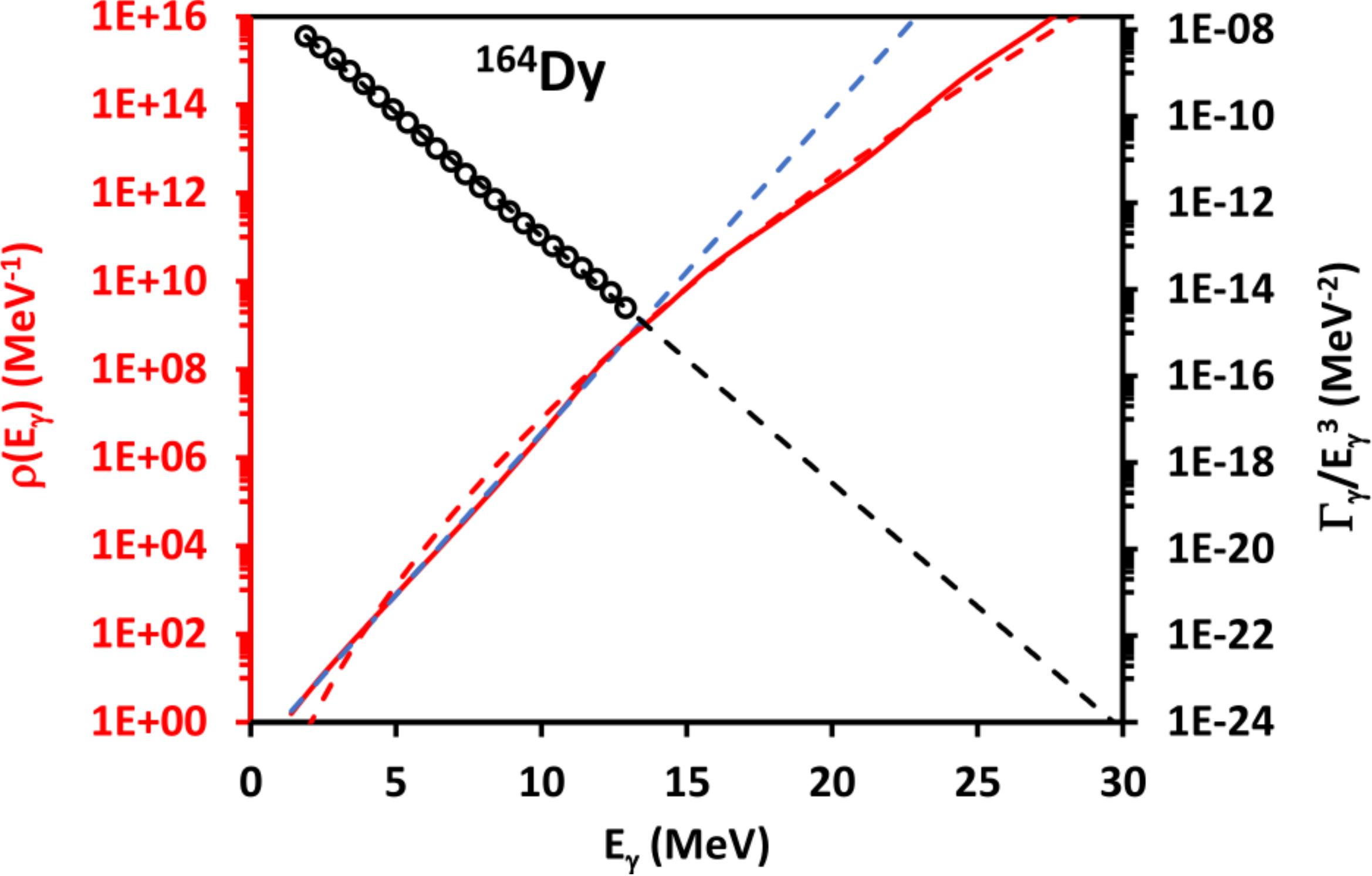}
\caption{Average reduced widths calculated from the CT-JPI and BA models below the GDR energy (\textbf{o}) and extrapolated to 30 MeV (\textbf{-~-~-}).  Level densities are from the CT-JPI model (\textcolor{blue}{\textbf{-~-~-}}), fitted to the BA model (\textcolor{red}{$\boldsymbol{-}$}), and calculated from the FGLD model (\textcolor{red}{\textbf{-~-~-}).}}
\end{figure*}
\setcounter{figure}{2}
\begin{figure*}[p]
\centering
\includegraphics[width=8.5cm]{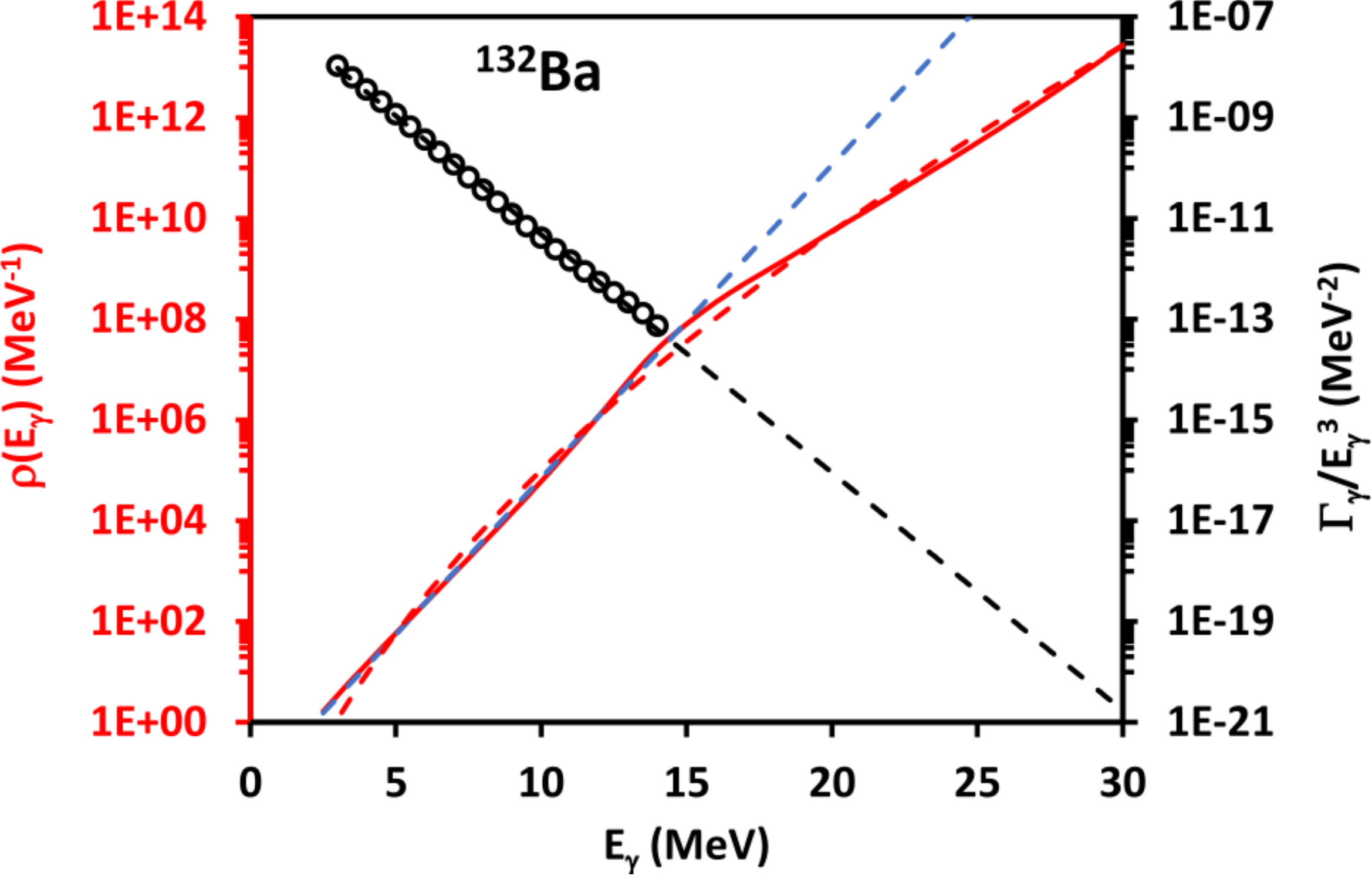}
\includegraphics[width=8.5cm]{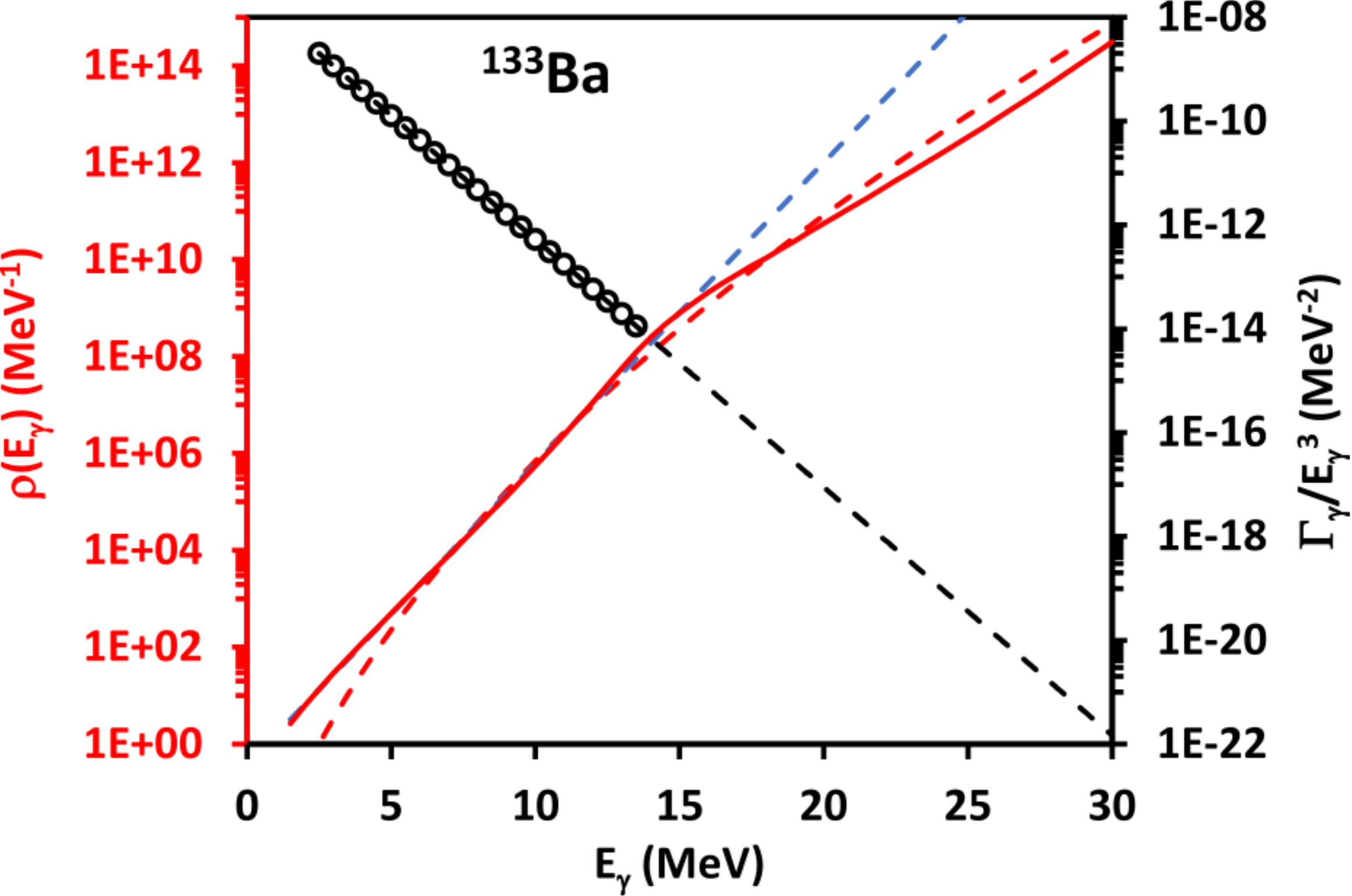}
\includegraphics[width=8.5cm]{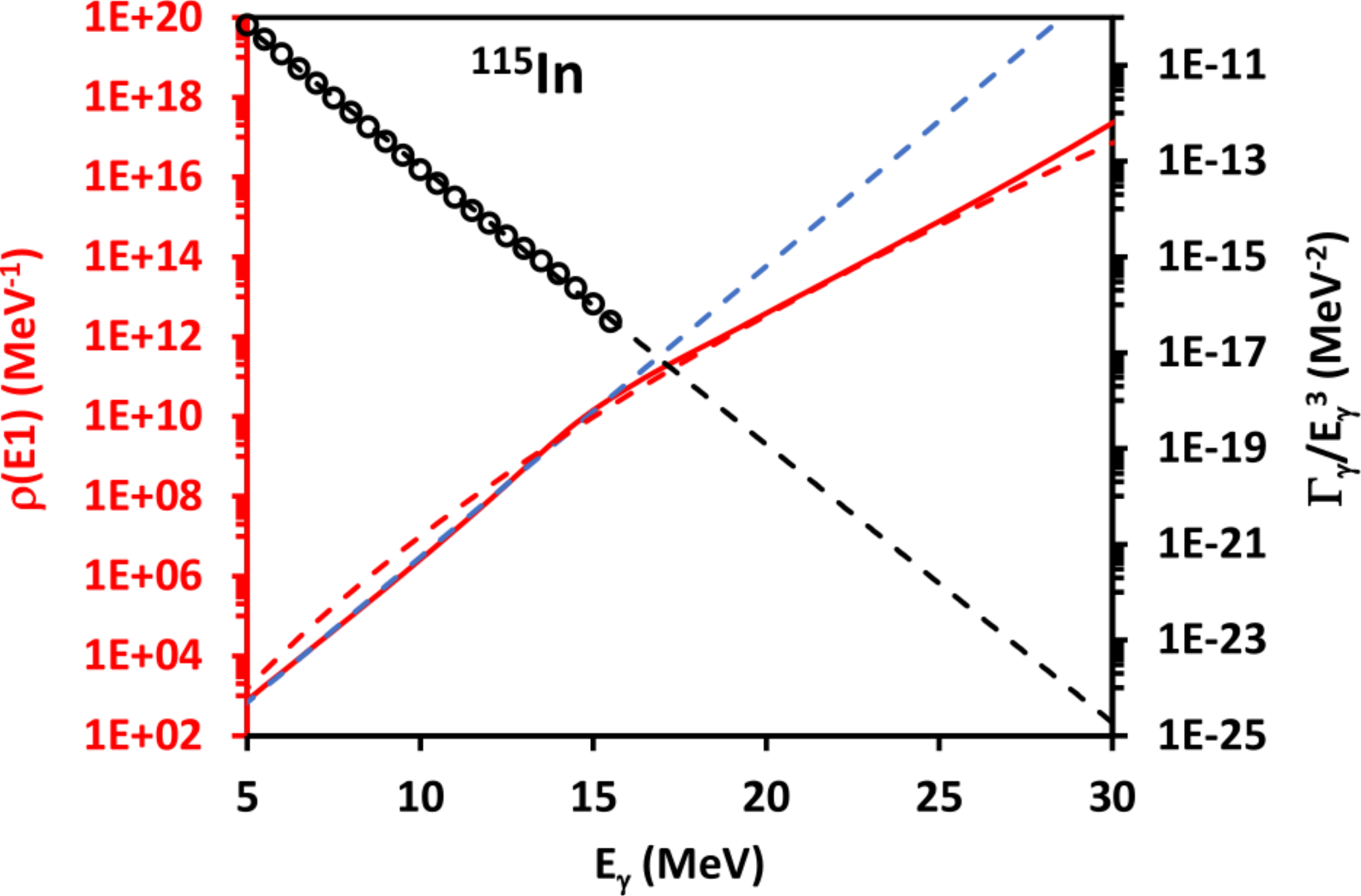}
\includegraphics[width=8.5cm]{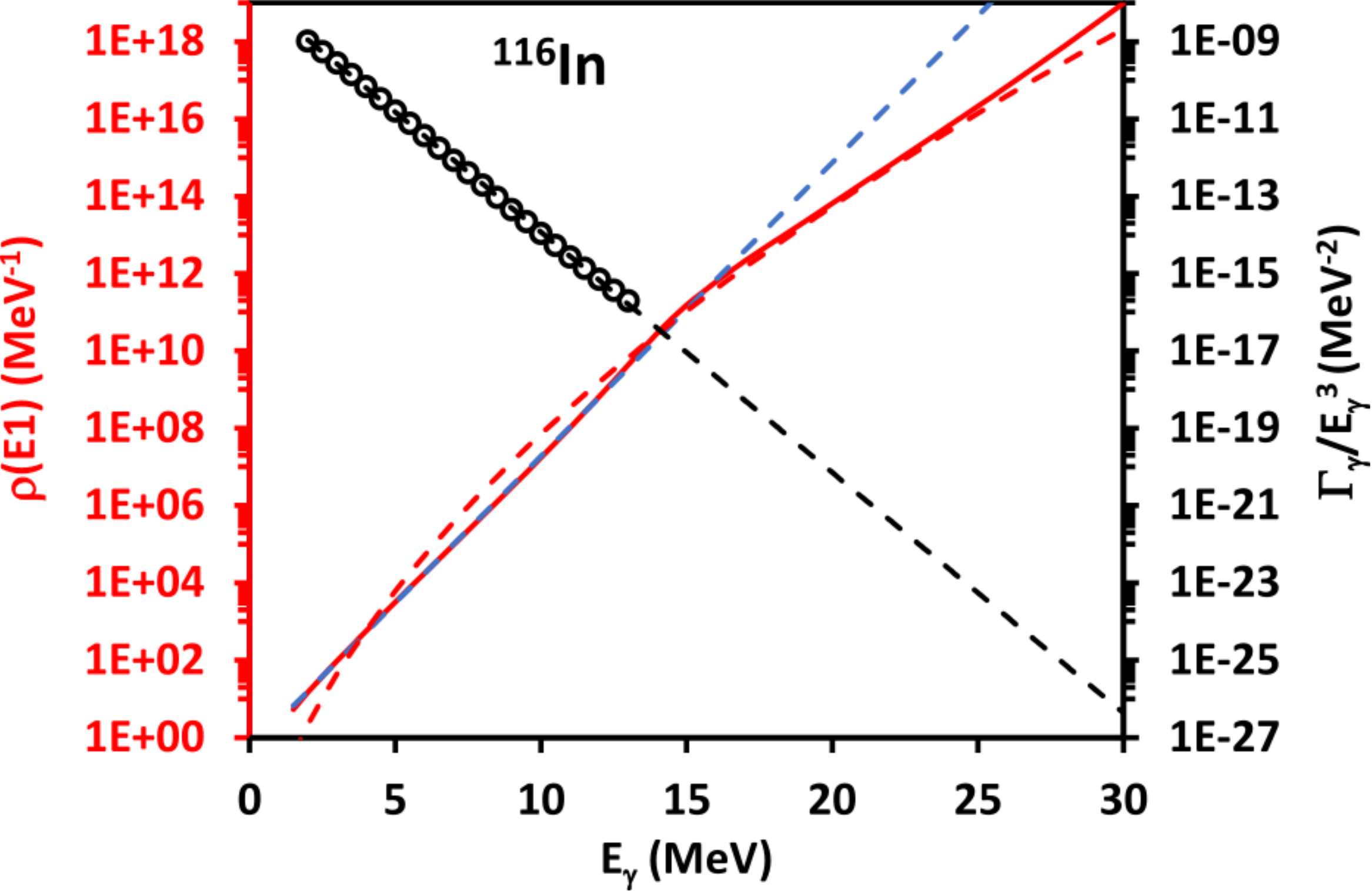}
\includegraphics[width=8.5cm]{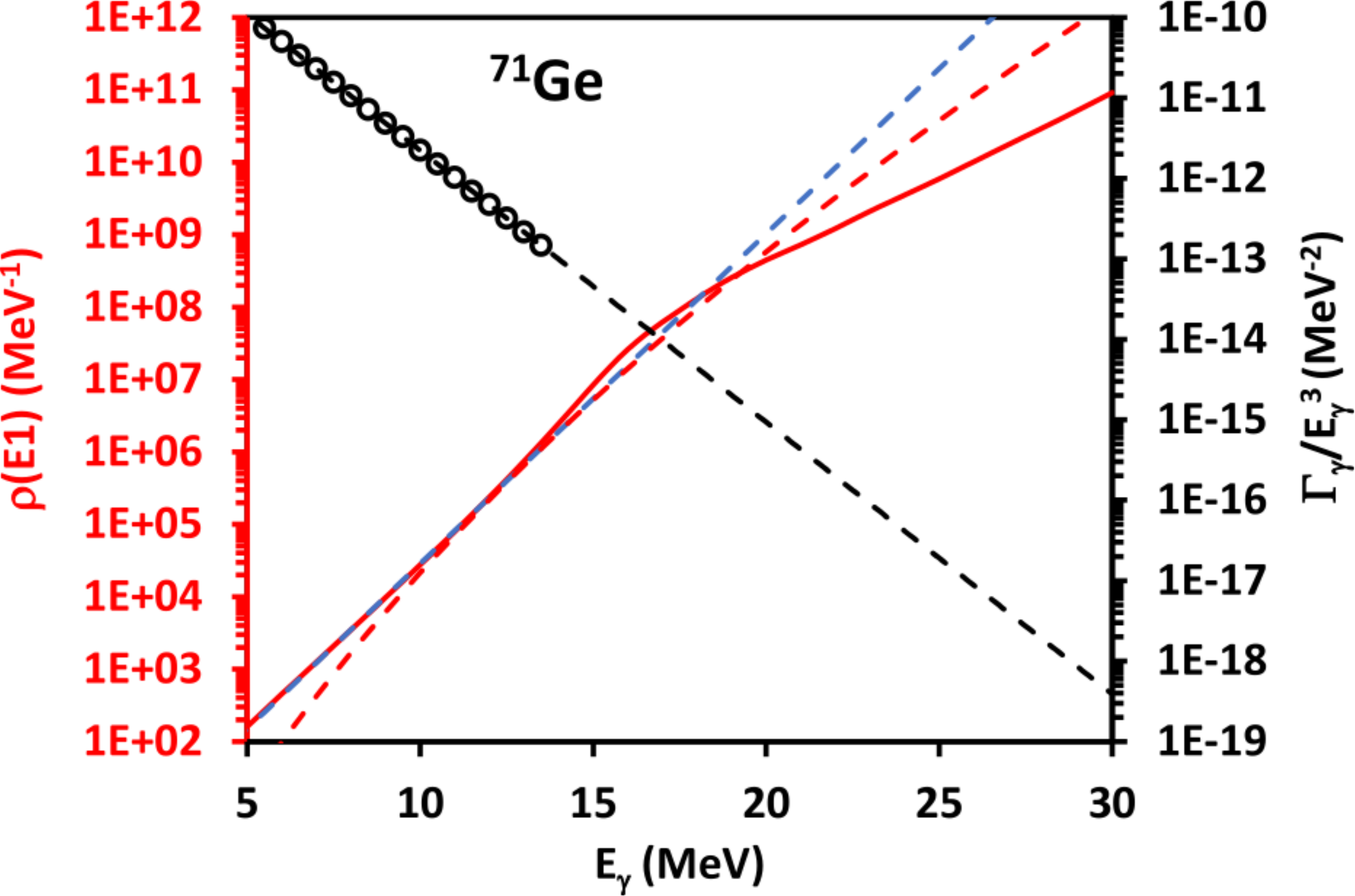}
\includegraphics[width=8.5cm]{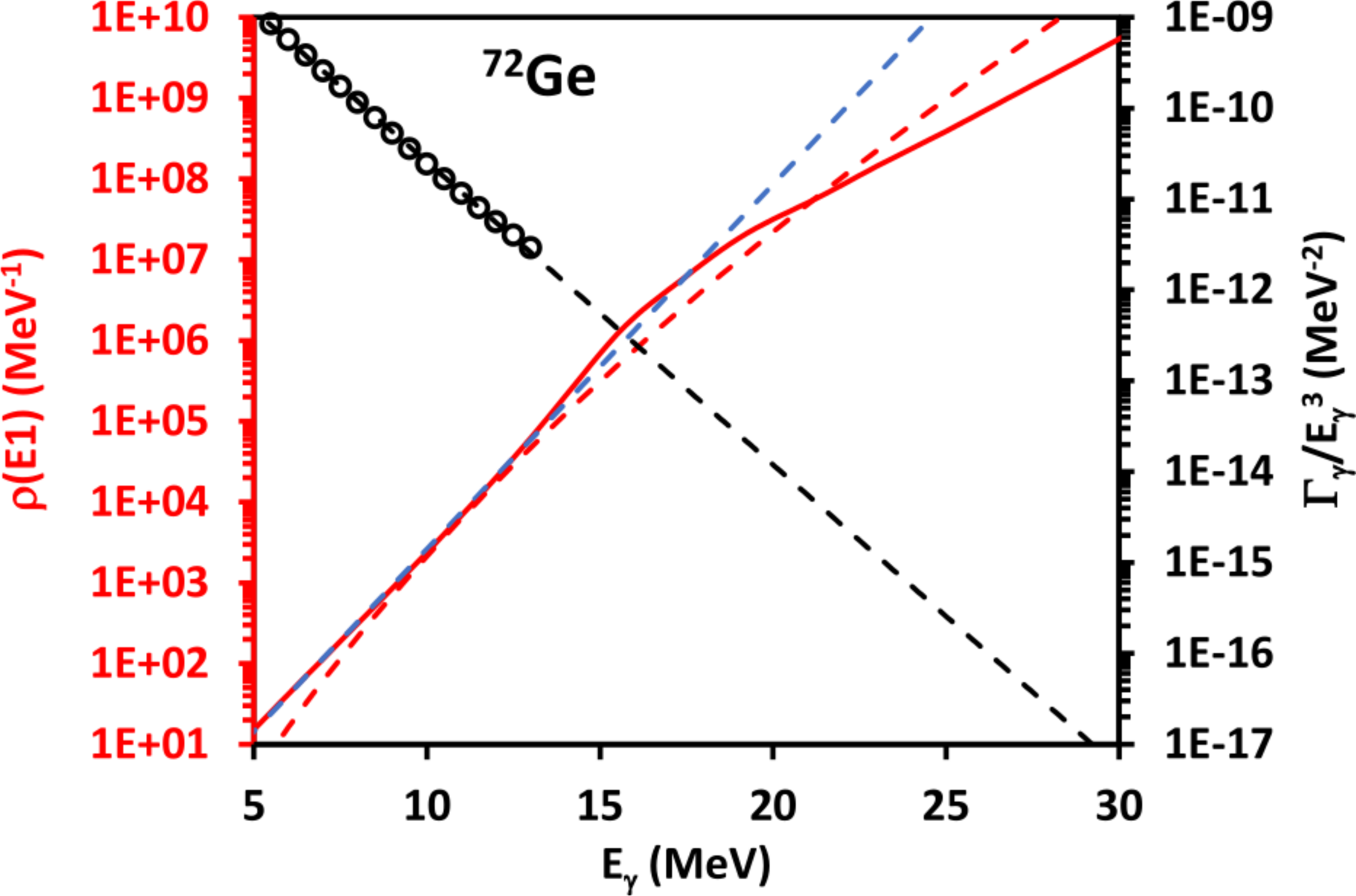}
\includegraphics[width=8.5cm]{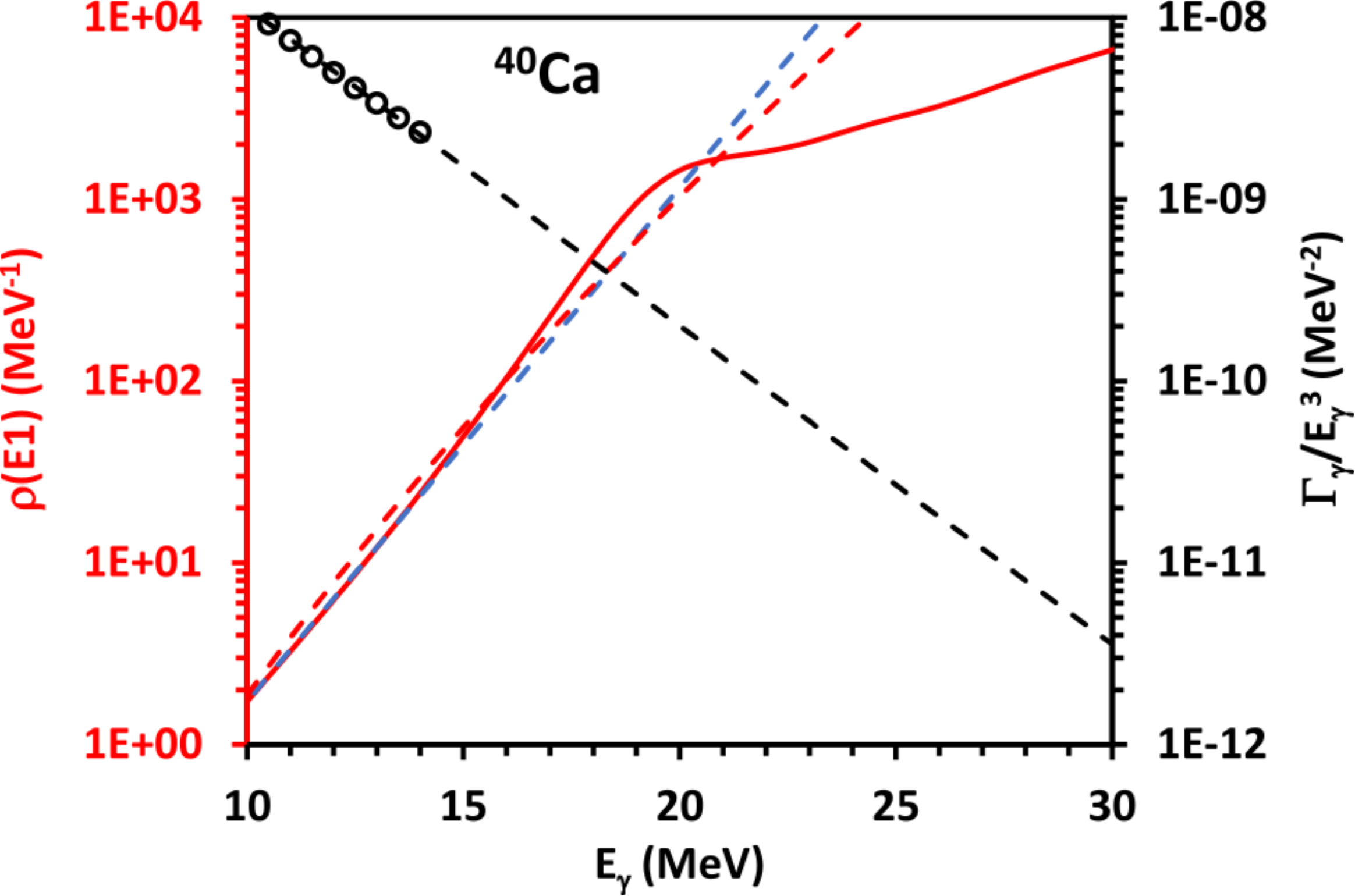}
\includegraphics[width=8.5cm]{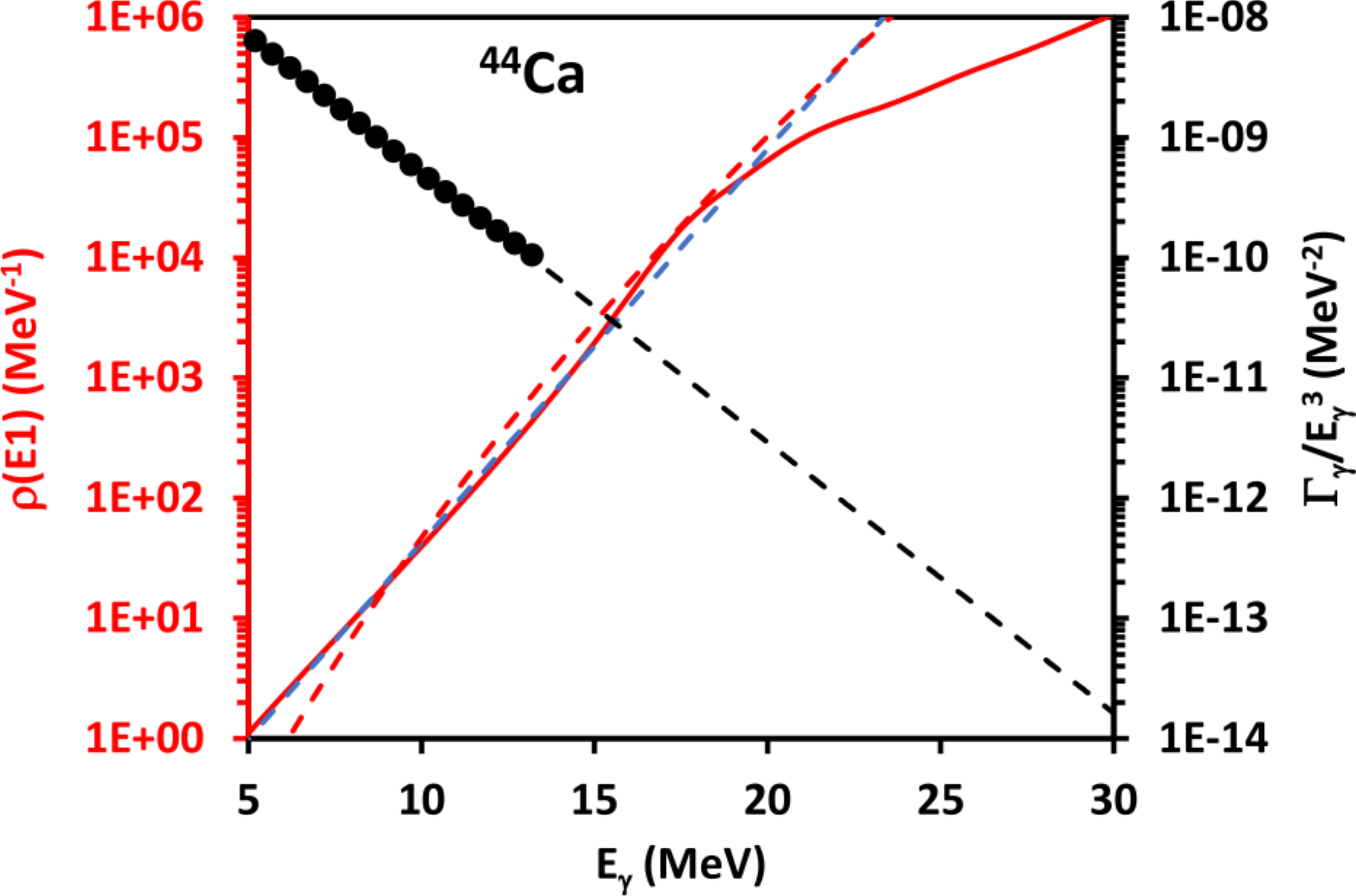}
\caption{\textbf{continued.}  Average reduced widths calculated from the CT-JPI and BA models below the GDR energy (\textbf{o}) and extrapolated to 30 MeV (\textbf{-~-~-}).  Level densities are from the CT-JPI model (\textcolor{blue}{\textbf{-~-~-}}), fitted to the BA model (\textcolor{red}{$\boldsymbol{-}$}), and calculated from the FGLD model (\textcolor{red}{\textbf{-~-~-}).}}
\label{Plot}
\end{figure*}

\begin{table}[!ht]
\tabcolsep=2pt
\caption{\label{TE0} CT-JPI model and fitted average reduced width parameters.}
\begin{tabular}{rcccccc}
\toprule
\multirow{2}{*}{$^A$El$_Z$}&\multirow{2}{*}{$J^{\pi}_i$}&\multirow{2}{*}{$J^{\pi}_f$}&$T$&$E_0$&\multicolumn{2}{c}{Fitting parameters$^a$}\\
&&&\multicolumn{2}{c}{MeV}&A&B\\
\colrule
\vspace{-0.3cm}\\
$^{238}$U$_{92}$&0$^+$&1$^-$&0.420&0.943&7.80$\times 10^{-8}$&-1.98\\
$^{235}$U$_{92}$&7/2$^-$&5/2$^+$&0.455&0.089&5.68$\times 10^{-9}$&-1.92\\
&&7/2$^+$&&0.113&&\\
&&9/2$^+$&&0.195&&\\
$^{208}$Pb$_{82}$&0$^+$&1$^-$&0.765&5.110&1.12$\times 10^{-5}$&-0.90\\
$^{207}$Pb$_{82}$&1/2$^-$&1/2$^+$&0.746&4.526&1.94$\times 10^{-6}$&-0.94\\
&&3/2$^+$&&3.886&&\\
$^{164}$Ho$_{67}$&1$^+$&0$^-$&0.559&0.999&1.02$\times 10^{-8}$&-1.42\\
&&1$^-$&&0.660&&\\
&&2$^-$&&0.537&&\\
$^{163}$Ho$_{67}$&7/2$^-$&5/2$^+$&0.535&0.339&6.30$\times 10^{-9}$&-1.50\\
&&7/2$^+$&&0.431&&\\
&&9/2$^+$&&0.588&&\\
$^{164}$Dy$_{66}$&0$^+$&1$^-$&0.593&1.367&8.64$\times 10^{-8}$&-1.31\\
$^{161}$Dy$_{66}$&5/2$^+$&3/2$^-$&0.548&0.493&5.13$\times 10^{-9}$&-1.46\\
&&5/2$^-$&&0.312&&\\
&&7/2$^-$&&0.212&&\\
$^{133}$Ba$_{56}$&1/2$^+$&1/2$^-$&0.696$^b$&1.284$^c$&3.11$\times 10^{-8}$&-1.10\\
&&3/2$^-$&0.696$^b$&1.583$^c$&&\\
$^{132}$Ba$_{56}$&0$^+$&1$^-$&0.699$^b$&2.453$^c$&2.48$\times 10^{-7}$&-1.09\\
$^{116}$In$_{49}$&1$^+$&0$^-$&0.572&1.600&1.94$\times 10^{-8}$&-1.43\\
&&1$^-$&&1.446&&\\
&&2$^-$&&1.140&&\\
$^{115}$In$_{49}$&9/2$^+$&7/2$^-$&0.596&1.648&4.89$\times 10^{-8}$&-1.34\\
&&9/2$^-$&&2.189&&\\
&&11/2$^-$&&3.306&&\\
$^{72}$Ge$_{32}$&0$^+$&1$^-$&0.962&2.470&5.23$\times 10^{-8}$&-0.76\\
$^{71}$Ge$_{32}$&1/2$^-$&1/2$^+$&0.953&1.308&5.34$\times 10^{-9}$&-0.78\\
&&3/2$^+$&&0.656&&\\
$^{44}$Ca$_{20}$&0$^+$&1$^-$&1.328&4.637&7.26$\times 10^{-8}$&-0.49\\
$^{40}$Ca$_{20}$&0$^+$&1$^-$&1.537&8.497&6.41$\times 10^{-7}$&-0.40\\
$^{28}$Si$_{14}$&0$^+$&1$^-$&1.518$^b$&8.905$^c$&4.60$\times 10^{-7}$&-0.43\\
\botrule
\vspace{-0.3cm}\\
\multicolumn{7}{l}{$^a (\Gamma_{\gamma}/E_{\gamma}^3)_{fit} =Ae^{BE_{\gamma}}$, $^b$Systematic value, $^c$Yrast energy.}\\
\end{tabular}
\end{table}

\section{Comparison with FGLD model}

\begin{figure}[!h]
\centering
\includegraphics[width=8cm]{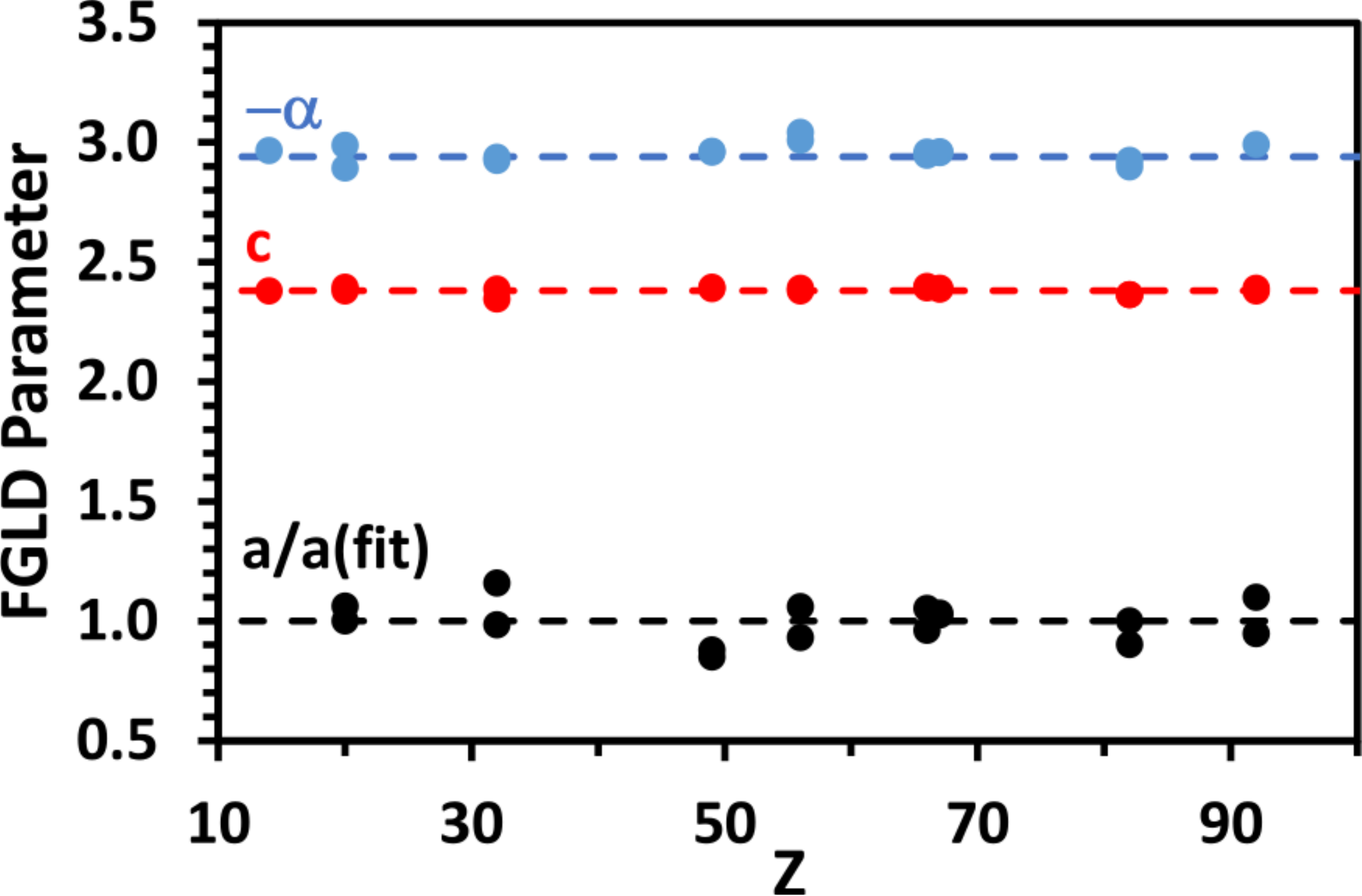}
\caption{Fitted FGLD model level density parameters, a, c, and the ratio of a to the value fitted to the equation $a=\frac{78(3)}{T^2 \sqrt{Z}}$.}
\label{LDA}
\end{figure}

Level densities calculated with the FGLD model are plotted in Fig.~\ref{Plot}.  The model parameters were least squares fit to the BA/CT-JPI level densities as shown in Eq.~\ref{FFG}.  The arbitrary logarithmic normalization
\begin{equation}\label{FFG}
 \sum_{E_i=E_0}^{30 MeV} \frac{ln\big(\rho(E_x)_i^{FGLD}\big)}{ln\big(\rho(E_x)_i^{BA/CTJPI}\big)}=1
\end{equation}
constrains it from being dominated by high energy values that tend to deviate most strongly.  The agreement between the two models is remarkable up to 30 MeV for all but the lightest nuclei.  The principle difference is that the FGLD model does not account for the level density increase at the shell closures.

The fitted FGLD model parameters for the nuclei investigated here are plotted in Fig.~\ref{LDA}.  The normalization parameter, $N_{FG}=2\times10^{-5}$, was found to give the best fit in all cases.  The contribution of the $\beta$ parameter is small and could be ignored.  The $\alpha$ parameter is nearly constant with an average value $\alpha$=2.383(14) which is in excellent agreement with $\alpha$=2.37(10) from Sen’kov and Zelevinsky~\cite{Senkov16} derived for $^{28}$Si.  The average constant value, $c=-2.96(4)$, is also in excellent agreement with $c=-2.92(13)$ by Sen’kov and Zelevinsky.  The level density parameter $a$ varied with atomic mass and could accurately be fit by either the relation $a=\frac{78(3)}{T^2}Z^{-1/2}$ or $a=\frac{119(3)}{T^2}A^{-1/2}$ although the dependence on A or Z cannot is undetermined.

\section{Discussion}

The BA formulation of photon strength is fit to experimental observations and its accuracy is well documented.  It makes no statement about the theoretical origin of the GDR because it is a peak fitting algorithm.  The strong correlation of the GDR energy with the $2\hbar\omega$ shell closure is evidence that the GDR peak is a shell effect consistent with the appearance of an ensemble of new levels.  This implies that similar resonances must exist at all shell closures.  The $1\hbar\omega$ shell closure is readily associated with the onset of pigmy and spin flip resonances.  All other resonances, i.e. GQR, GMR, GOR, ISGDR, are observed to occur at $2\hbar\omega-4\hbar\omega$ shell closures~\cite{Firestone20}.  The extension of the BA formulation to include higher energy resonances is natural and necessary.

The CT-JPI level density model is also based on experimental observations.  It has been shown~\cite{Firestoneb21} that level density varies exponentially for each spin and parity defined by two parameters, a backshift energy $E_0(J^{\pi})$ and a temperature $T$.  These parameters are constrained by experimental level energies and a statistical spin distribution function based on a single spin cutoff parameter.  The CT-JPI model provides experimental level densities for levels below the GDR energy that are populated by E1 transitions in photonuclear reactions.

Photon strength is separable into experimentally determined level density and reduced level width components.  Deconvolution of the BA formulation with the CT-JPI model level densities determines the reduced level widths at low energies.  Extrapolating the reduced level widths exponentially to higher energies bootstraps the use of the BA formulation to determine level densities at higher energies.  The remaining uncertainty is whether the exponential extrapolation of the reduced level widths to high energies is valid.  In all cases the statistical uncertainty in the exponential fits is $\approx$2\% with a coefficient of determination $R^2$=0.99.  The extrapolation only applies to the partial reduced level width associated with the primary $(\gamma,n)$ or $(\gamma,f)$ reaction so as additional reaction channels open the total reduced level width will increase accordingly.

The assumption that the photonuclear reaction is dominated by E1 transition multipolarity also is uncertain.  M1 transitions follow the same $E_{\gamma}^3$ energy dependence as E1 transitions and are unlikely to become important and higher energies.  E2 transitions follow an $E_{\gamma}^5$ energy dependence so even if their contribution were small at low energies they could became significant at high energies.  A correction for E2 transitions would increase both the reduced average level width and the calculated level density as levels with $J^{\pi}=J_{GS}^{\pi},J_{GS}^{\pi}\pm1,J_{GS}^{\pi}\pm2$ become populated.

The qualitative agreement of the calculated level densities with the FGLD model is theoretical confirmation that the BA/CT-JPI analysis is correct.  This is not surprising as this approach mirrors that of Gilbert and Cameron~\cite{Gilbert65} who proposed combining the Constant Temperature (CT) model at low energies with the BSFG model at higher energies, matching the two at the neutron separation energy.  Their approach was flawed because the CT model used to calculate the total level density leads to nonphysical fitting parameters and no $J^{\pi}$ dependence.  The CT-JPI model corrects this deficiency.

\begin{figure}[!h]
  \centering
\includegraphics[width=8cm]{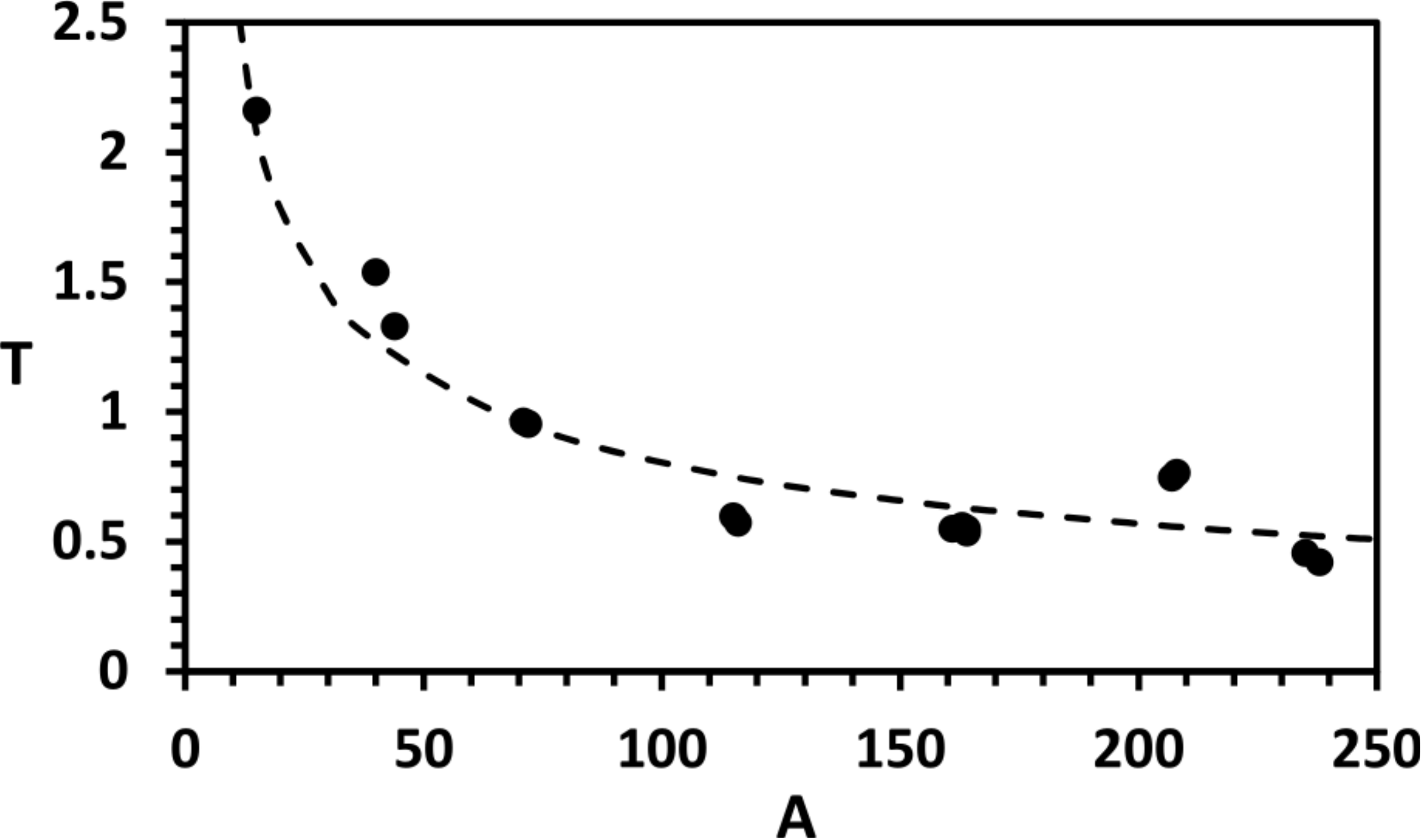}
\caption{Temperatures of selected nuclei~\cite{Firestoneb21} ($\bullet$) fit to function $T=0.80(4)/\sqrt{A}$ (\textbf{-~-~-}).}
\label{Temp}
\end{figure}

The FGLD model provides a quick method for estimating level densities, as shown in Eq.~14.  It only requires  \begin{equation*}
\begin{aligned}
\rho(E_{\gamma})\!=\!\frac{1}{E_{\gamma}^{5/4}}\textrm{exp}&\!\bigg[\frac{17.6(4)}{T}Z^{1/4}\!-\!2.96(4)\!-\!0.210(3)E^{-1/2}\bigg]\\
&\textrm{or}\hspace{5.1cm}(14)\\
\rho(E_{\gamma})\!=\!\frac{1}{E_{\gamma}^{5/4}}\textrm{exp}&\!\bigg[\frac{21.6(8)}{T}A^{1/4}\!-\!2.96(4)\!-\!0.210(3)E^{-1/2}\bigg]
\end{aligned}
\end{equation*}
a single parameter, $T$ although it fails to account the for the contribution of the GDR.  The temperatures of selected nuclei are plotted versus mass in Fig~\ref{Temp}.  They follow the empirical relationship $T=0.80(4)/\sqrt{A}$ with $\approx$15\% accuracy and significant deviations near closed shells.  The FGLD model can be used to calculate the level density for nuclei where little experimental data are available.

\begin{figure}[!h]
\centering
  \includegraphics[width=8.5cm]{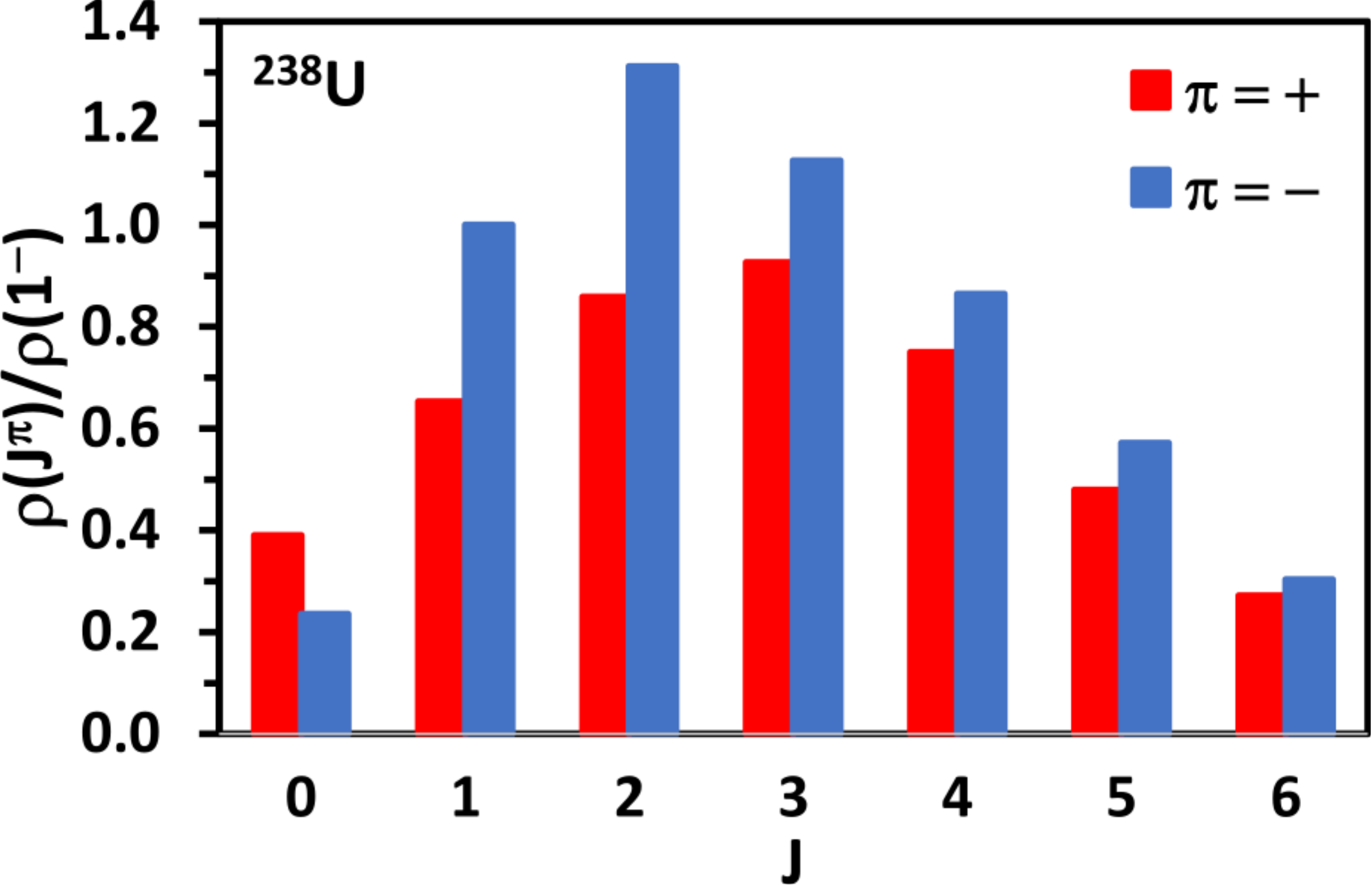}
\caption{Ratio of level densities $\rho(J^{\pi})/\rho(1^-)$ for $^{238}$U calculated with the CT-JPI model.}
\label{JPI}
\end{figure}

\setcounter{equation}{14}
This level densities described here are for levels populated by E1 transitions in photonuclear reactions.  The level densities for other $J^{\pi}$ values can be calculated with the CT-JPI model which is fitted to the energy independent spin distribution function~\cite{Eric60}.  For example, the ratios of level densities $J^{\pi}/J^{1^-}$ for $^{238}$U are given in Fig.~\ref{JPI}.  The energy independence of this ratio arises from the constant temperature for all spins and parities which causes their level densities to increase in parallel.

\section{Acknowledgements}

This work was supported by funding from the University of California retirement system.  Although no federal funding was provided I acknowledge the U.S. Department of Energy, Office of Nuclear Science, Nuclear Data Program for their long history of supporting my research that made this work possible.
%\clearpage
\bibliographystyle{apa}
%\bibliography{PS-BiB}	

\begin{thebibliography}{}

\bibitem[\protect\astroncite{Axel}{1962}]{Axel62}
Axel, P. (1962).
\newblock Electric dipole ground-state transition width strength function and
  7-mev photon interactions.
\newblock {\em Phys. Rev.}, 126:671--683.

\bibitem[\protect\astroncite{Bethe}{1936}]{Bethe36}
Bethe, H.~A. (1936).
\newblock An attempt to calculate the number of energy levels of a heavy
  nucleus.
\newblock {\em Phys. Rev.}, 50:332--341.

\bibitem[\protect\astroncite{Bethe}{1937}]{Bethe37}
Bethe, H.~A. (1937).
\newblock Nuclear physics b. nuclear dynamics, theoretical.
\newblock {\em Rev. Mod. Phys.}, 9:69--244.

\bibitem[\protect\astroncite{Brink}{1955}]{Brink55}
Brink, D. (1955).
\newblock {\em Some aspects of the interaction of light with matter}.
\newblock PhD thesis, University of Oxford.

\bibitem[\protect\astroncite{Capote et~al.}{2009}]{Capote09}
Capote, R., Herman, M., Obložinský, P., Young, P., Goriely, S., Belgya, T.,
  Ignatyuk, A., Koning, A., Hilaire, S., Plujko, V., Avrigeanu, M., Bersillon,
  O., Chadwick, M., Fukahori, T., Ge, Z., Han, Y., Kailas, S., Kopecky, J.,
  Maslov, V., Reffo, G., Sin, M., Soukhovitskii, E., and Talou, P. (2009).
\newblock Ripl – reference input parameter library for calculation of nuclear
  reactions and nuclear data evaluations.
\newblock {\em Nuclear Data Sheets}, 110(12):3107--3214.
\newblock Special Issue on Nuclear Reaction Data.

\bibitem[\protect\astroncite{Chadwick et~al.}{2000}]{IAEA1178}
Chadwick, M., Blokhin, A., Fukahori, T., Lee, Y.-O., Martins, M., Varmalov, V.,
  Yu, B., Han, Y., Mughabghab, S., and Zhang, J. (2000).
\newblock {\em Handbook on Photonuclear Data for Applications Cross-sections
  and Spectra}.
\newblock Number 1178 in TECDOC Series. INTERNATIONAL ATOMIC ENERGY AGENCY,
  Vienna.

\bibitem[\protect\astroncite{Ericson}{1960}]{Eric60}
Ericson, T. (1960).
\newblock The statistical model and nuclear level densities.
\newblock {\em Advances in Physics}, 9(36):425--511.

\bibitem[\protect\astroncite{{Firestone}}{2020}]{Firestone20}
{Firestone}, R.~B. (2020).
\newblock {The Origin of the Giant Dipole Resonance}.
\newblock {\em arXiv e-prints}, page arXiv:2009.03356.

\bibitem[\protect\astroncite{{Firestone}}{2021}]{Firestoneb21}
{Firestone}, R.~B. (2021).
\newblock {Spin/Parity Dependent Level Density}.
\newblock {\em arXiv e-prints}, page arXiv:2104.02693.

\bibitem[\protect\astroncite{{Gilbert} and {Cameron}}{1965}]{Gilbert65}
{Gilbert}, A. and {Cameron}, A.~G.~W. (1965).
\newblock {A composite nuclear-level density formula with shell corrections}.
\newblock {\em Canadian Journal of Physics}, 43:1446.

\bibitem[\protect\astroncite{Goldhaber and Teller}{1948}]{Goldhaber46}
Goldhaber, M. and Teller, E. (1948).
\newblock On nuclear dipole vibrations.
\newblock {\em Phys. Rev.}, 74:1046--1049.

\bibitem[\protect\astroncite{Goriely et~al.}{2008}]{Goriely08}
Goriely, S., Hilaire, S., and Koning, A.~J. (2008).
\newblock Improved microscopic nuclear level densities within the
  hartree-fock-bogoliubov plus combinatorial method.
\newblock {\em Phys. Rev. C}, 78:064307.

\bibitem[\protect\astroncite{Goriely et~al.}{2007}]{Goriely07}
Goriely, S., Samyn, M., and Pearson, J.~M. (2007).
\newblock Further explorations of skyrme-hartree-fock-bogoliubov mass formulas.
  vii. simultaneous fits to masses and fission barriers.
\newblock {\em Phys. Rev. C}, 75:064312.

\bibitem[\protect\astroncite{Kawano et~al.}{2020}]{Kawano19}
Kawano, T., Cho, Y., Dimitriou, P., Filipescu, D., Iwamoto, N., Plujko, V.,
  Tao, X., Utsunomiya, H., Varlamov, V., Xu, R., Capote, R., Gheorghe, I.,
  Gorbachenko, O., Jin, Y., Renstrøm, T., Sin, M., Stopani, K., Tian, Y.,
  Tveten, G., Wang, J., Belgya, T., Firestone, R., Goriely, S., Kopecky, J.,
  Krtička, M., Schwengner, R., Siem, S., and Wiedeking, M. (2020).
\newblock Iaea photonuclear data library 2019.
\newblock {\em Nuclear Data Sheets}, 163:109 -- 162.

\bibitem[\protect\astroncite{Kopecky and Uhl}{1990}]{Kopecky90}
Kopecky, J. and Uhl, M. (1990).
\newblock Test of gamma-ray strength functions in nuclear reaction model
  calculations.
\newblock {\em Phys. Rev. C}, 41:1941--1955.

\bibitem[\protect\astroncite{Möller et~al.}{2016}]{Moller16}
Möller, P., Sierk, A., Ichikawa, T., and Sagawa, H. (2016).
\newblock Nuclear ground-state masses and deformations: Frdm(2012).
\newblock {\em Atomic Data and Nuclear Data Tables}, 109-110:1–204.

\bibitem[\protect\astroncite{Newton}{1956}]{Newton56}
Newton, T.~D. (1956).
\newblock Shell effects on the spacing of nuclear levels.
\newblock {\em Canadian Journal of Physics}, 34(8):804--829.

\bibitem[\protect\astroncite{{Pritychenko} et~al.}{2014}]{Pritychenko14}
{Pritychenko}, B., {Birch}, M., {Horoi}, M., and {Singh}, B. (2014).
\newblock {B(E2) Evaluation for {}0$_{1}$$^{+}$ {\textrightarrow}
  {}2$_{1}$$^{+}$ Transitions in Even-Even Nuclei}.
\newblock {\em Nuclear Data Sheets}, 120:112--114.

\bibitem[\protect\astroncite{Sen'kov and Zelevinsky}{2016}]{Senkov16}
Sen'kov, R. and Zelevinsky, V. (2016).
\newblock Nuclear level density: Shell-model approach.
\newblock {\em Phys. Rev. C}, 93:064304.

\bibitem[\protect\astroncite{Uhl and Kopecky}{1995}]{Uhl94}
Uhl, M. and Kopecky, J. (1995).
\newblock Gamma-ray strength function models and their parameterization.
\newblock Technical Report ECN-RX-94-099, Netherlands.

\end{thebibliography}

\end{document}